\documentclass[11pt,a4paper,preprint,amsmath,amssymb,tightenlines]{article}
\pdfoutput=1
\usepackage{jheppub}
\usepackage[english]{babel}
\usepackage{graphicx}
\usepackage{bm}
\usepackage{subcaption}
\usepackage{slashed}

\usepackage{xspace}
\usepackage{hyperref}
\usepackage[usenames,dvipsnames]{xcolor}
\usepackage{feynmp-auto}
\graphicspath{{figs/}}

\allowdisplaybreaks[4]

\newcommand{\beq}{\begin{equation}\begin{aligned}}
\newcommand{\eeq}{\end{aligned}\end{equation}}

\def\Li{{\rm Li}}
\def\cN{{\mathcal N}}
\def\cM{{\mathcal M}}
\def\cO{{\mathcal O}}

\def\re{\text{Re}}
\def\im{\text{Im}}

\definecolor{darkyellow}{rgb}{0.5, 0.5, 0.0}
\definecolor{darkpurple}{rgb}{0.5, 0.2, 0.8}
\definecolor{darkblue}{rgb}{0.0, 0.0, 0.8}
\definecolor{darkgreen}{rgb}{0.0, 0.4, 0.0}
\definecolor{darkred}{rgb}{0.5, 0.0, 0.0}

\hypersetup{
    linktocpage,
     colorlinks,
     citecolor=darkgreen,
     linkcolor= darkgreen,
     urlcolor=darkgreen
}

\newcommand{\red}[1]{ {\color{darkred}{#1}}}
\newcommand{\green}[1]{ {\color{darkgreen}{#1}}}

\preprint{MSUHEP-22-025, ZU-TH 38/22}
\title{Analytic Computation of Three-point Energy Correlator in QCD}

\author[1,2]{Tong-Zhi Yang,}
\author[3]{Xiaoyuan Zhang}

\affiliation[1]{Physik-Institut, Universit\"at Z\"urich, Winterthurerstrasse 190, CH-8057 Z\"urich, Switzerland}
\affiliation[2]{Department of Physics and Astronomy, Michigan State University, 48824, East Lansing, MI, US}
\affiliation[3]{Department of Physics, Harvard University, 02138, Cambridge, MA, US}

\emailAdd{toyang@physik.uzh.ch}
\emailAdd{xiaoyuanzhang@g.harvard.edu}

\abstract{
The energy correlator measures the energy deposited in multiple detectors as a function of the angles among them. In this paper, an analytic formula is given for the three-point energy correlator with full angle dependence at leading order in electron-positron annihilation. This is the first analytic computation of trijet event shape observables in QCD, which provides valuable data for phenomenological studies.
The result is computed with direct integration, where appropriate parameterizations of both phase space and kinematic space are adopted to simplify the calculation.
With full shape dependence, our result provides the expansions in various kinematic regions such as equilateral, triple collinear and squeezed limits, which benefit 
studies on both factorization and large logarithm resummation.
}

\begin{document}
\maketitle
\flushbottom
\newpage

\section{Introduction}
One of the event shape observables that attracts lots of recent interest in quantum chromodynamics (QCD) and the collider physics community is \textit{energy correlators}. Traditionally, energy-energy correlation (EEC) measures the energy deposited in two detectors as a function of the angle between these two detectors \cite{Basham:1978zq,Basham:1978bw}. As observed in \cite{Basham:1978zq}, the fact that energy weights suppress the soft divergence makes EEC less sensitive to soft gluon emissions. More recently, EEC is generalized to a broader class of observables called energy correlators. In particular, the three-point energy correlator (EEEC), which depends on the three angles among the detectors, contains the nontrivial shape information of the scattering process \cite{Chen:2019bpb,Chen:2020vvp, Yan:2022cye}. In perturbative theories, EEEC is defined as
\begin{multline}\label{eq:eeecdef}
\frac{1}{\sigma_{\text{tot}}}\frac{d^3\sigma}{dx_1dx_2dx_3}= \sum_{ijk}\int d\sigma \frac{E_iE_jE_k}{Q^3}\\
\times\delta\left(x_1-\frac{1-\cos\theta_{jk}}{2}\right) \delta\left(x_2-\frac{1-\cos\theta_{ik}}{2}\right) \delta\left(x_3-\frac{1-\cos\theta_{ij}}{2}\right)\,,
\end{multline}
where $i$, $j$ and $k$ run over all final-state particles, $Q$ is the total energy of the electron-positron annihilation, and $d\sigma$ is the differential cross section. For convenience, we normalize the distribution to the born cross section. EEEC is infrared finite in the tree-level $\gamma^{*}\to 4 \text{ jets}$ process, which allows us to perform the calculation in $d=4$ dimension.

\begin{figure}[!htp]
	\centering
	\begin{subfigure}{.4\linewidth}
		\includegraphics[scale=0.3]{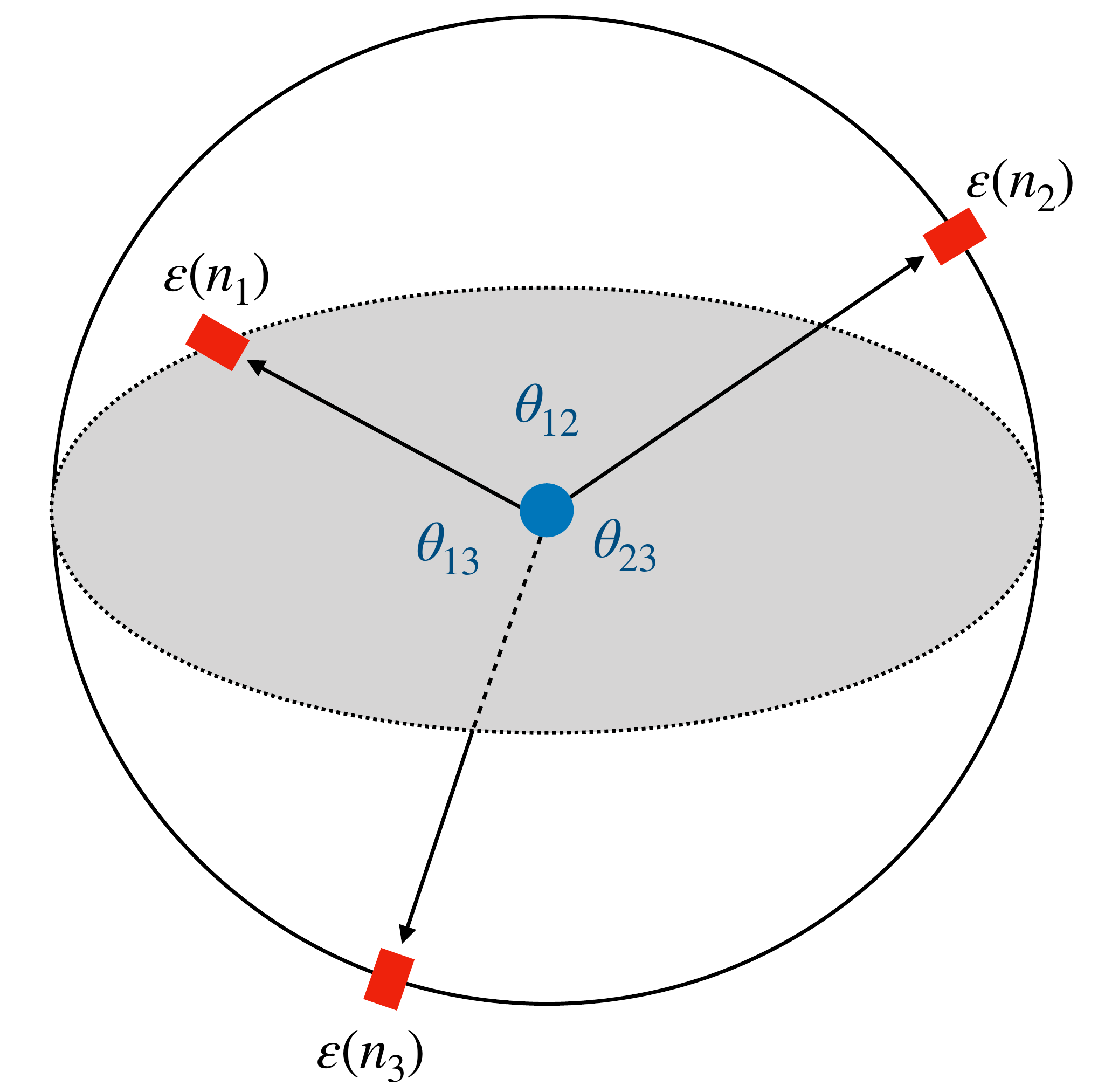}
		\caption{}
		\label{fig:eeec_picture}
	\end{subfigure}
	\quad\quad
	\begin{subfigure}{.4\linewidth}
		\includegraphics[scale=0.9]{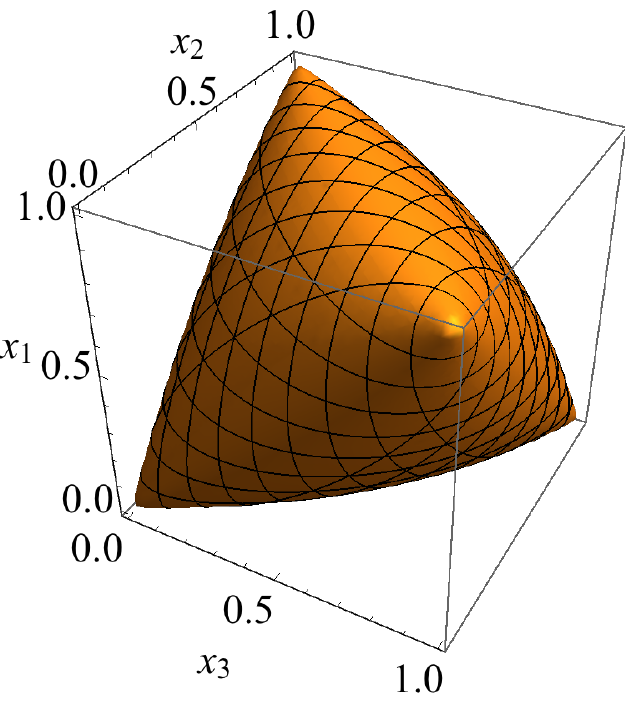}
		\caption{}
		\label{fig:kinematic_region}
	\end{subfigure}
	\caption{(a) A graph on the three-point energy correlator. The three detectors are separated by finite angles $\theta_{12}$, $\theta_{13}$ and $\theta_{23}$, capturing outgoing particles at specific angles from the hard interaction and summing their energies. (b) The ``zongzi''-shaped kinematic space $\{x_1,x_2,x_3\}$, which is constrained by the four-particle phase space.}
	\label{fig:eeec_plot}
\end{figure}
Energy correlators are \textit{almost} the simplest infrared (IR) safe jet observables to compute analytically. The leading order (LO) EEC in QCD is obtained since 1970s \cite{Basham:1978zq,Basham:1978bw}. Recently, EEC is also computed analytically to next-to-leading order (NLO) in QCD \cite{Dixon:2018qgp,Luo:2019nig,Gao:2020vyx} and NNLO in $\cN=4$ super Yang-Mills (SYM) theory \cite{Belitsky:2013ofa,Henn:2019gkr}. At the same time, the collinear limit of the LO EEEC in both $\cN=4$ SYM and QCD is studied in \cite{Chen:2019bpb} and the complete LO $\cN=4$ SYM result becomes available very recently \cite{Yan:2022cye}. In this paper, we calculate the complete LO EEEC in QCD, which shares a similar function space and analytic structure as in $\cN=4$ SYM. 

There is also lots of progress in studying energy correlators with effective field theories (EFTs), such as Soft-Collinear Effective theory (SCET) \cite{Bauer:2000yr,Bauer:2000ew,Bauer:2001yt,Bauer:2001ct,Beneke:2002ph}, which proves to be essential in jet substructure. As summarized in \cite{Dixon:2018qgp}, EEC is both singular in the collinear and back-to-back limits, and large logarithms in both limits could possibly spoil the perturbation theory. Regarding the collinear region, the resummation has been achieved to the next-to-next-to-leading logarithm (NNLL) accuracy in QCD~\cite{Dixon:2019uzg} and $\cN=4$ SYM~\cite{Korchemsky:2019nzm}. In the back-to-back limit, EEC is resummed to NNLL accuracy and matched to NNLO fixed-order prediction \cite{COLLINS1981381,ELLIS198499,deFlorian:2004mp,Tulipant:2017ybb,Moult:2019vou}, while a new factorization formula is also introduced in \cite{Moult:2018jzp}, allowing the resummation to N$^3$LL~\cite{Ebert:2020sfi}. With the recently derived four loop rapidity anomalous dimension in QCD~\cite{Moult:2022xzt,Duhr:2022yyp}, EEC is also resummed to N$^4$LL accuracy in the back-to-back limit~\cite{Duhr:2022yyp}. EEC can also be studied at a hadron collider, the simplicity of the soft function allows the NNLL resummation in the back-to-back limit \cite{Gao:2019ojf}. More interestingly, the collinear factorization can also be generalized to EEEC observable, where the distribution is factorized into the convolution of a hard function and a jet function. While the factorization is straightforward in SCET, the resummation becomes subtle due to multiple variables. One way is to project the full kinematic region into a one-dimension space, which is referred to as the \textit{projected energy correlators} \cite{Chen:2020vvp}. The projected $N$-point correlator is defined as
\begin{equation}\label{eq:pecdef}
\frac{d\sigma}{dx_L}=\sum_n\sum_{1\leq i_1,\cdots i_N\leq n}\int d\sigma \frac{\prod_{a=1}^N E_{i_a}}{Q^N}\delta(x_L-\text{max}\{x_{i_1,i_2},x_{i_1,i_3}, \cdots x_{i_{N-1}, i_N}\})\,,
\end{equation}
and its collinear logarithms can be resummed to NNLL accuracy \cite{e3cnnll}. It would be also interesting to study EEEC in other kinematic limits. Since the shape dependence of EEEC provides more information on the jet substructure, several singular regions besides collinear remain unexplored: equilateral limit ($x_{1,2,3} \sim \eta$), squeezed limit ($x_1\sim 0, x_{2,3}\sim x$), coplanar limit and so on. Our fixed-order calculation allows one to extract both leading power (LP) and next-to-leading power (NLP) expansions, which benefit the large logarithm resummations. In a word, the energy correlator is a bridge to precision standard model tests and new physics searches.

The energy correlators attract lots of attention on the phenomenological side these days. In Ref.~\cite{Komiske:2022enw}, both the shape dependence and the scaling behavior of EEEC, as well as the ratio of projected energy correlators with respect to EEC are measured with the CMS open data. The close agreement between theoretical prediction and CMS open data proves that energy correlators will play an important role in precision QCD measurement and jet substructure, and it would be interesting to perform the measurement at the Large Hadron Collider (LHC). Besides, energy correlators enable measurements in hadronic environments to be theoretically predicted by means of modern loop computation techniques and track functions \cite{Chen:2020vvp,Li:2021zcf,Jaarsma:2022kdd}. Traditionally, the calculation of track-based observables  (e.g. angularities) requires the full functional form of track functions $T(x)$ \cite{Chang:2013rca}, of which the renormalization group evolution is described by complicated nonlinear equations. However, it is found recently that energy correlator is advantageous for studying track information since it only needs a finite number of track functions moments, which are just numbers and hence do not take part in the phase space integration. It is also suggested that an energy correlator can be applied to top quark mass measurement at the LHC \cite{Holguin:2022epo}.

It has been observed that $N$-point energy correlators can be written as $(N+2)$-point Wightman correlation function of energy flux operators and source operators that produces the localized excitation \cite{Hofman:2008ar}. Explicitly, EEEC can be alternatively defined by
\begin{multline}
\frac{d^3\sigma}{dx_1dx_2dx_3}\propto \int \prod_{i=1}^{3} \left[d\Omega_{\vec n_i} \delta\left(\frac{1-\vec n_i\cdot \vec n_{i+1}}{2}-x_i\right)\right]\\
\times \frac{\int d^4 x e^{iq \cdot x} \langle 0 |\cO^\dagger (x){\cal E}(\vec n_1){\cal E}(\vec n_2) {\cal E}(\vec n_3)\cO(0)|0\rangle }{Q^3\int d^4 x e^{iq\cdot x} \langle 0| \cO^\dagger (x) \cO(0)|0\rangle}\,,
\end{multline}
where the energy flux operator is given by integrated stress-energy tensor $T_{\mu\nu}$ along the direction $\vec n_i$ \cite{Sveshnikov:1995vi, Korchemsky:1997sy,Korchemsky:1999kt,Belitsky:2001ij}:
\begin{equation}
{\cal E}(\vec n)=\int_{-\infty}^{\infty} d\tau \lim_{r\to \infty}r^2 n^i T_{0i}(t=\tau+r,r\vec n)\,,
\end{equation}
and for electron-positron collision, the source operator $\cO$ is the electromagnetic current. In conformal field theory (CFT), the light-ray operator product expansion (OPE) \cite{Kologlu:2019mfz,Chang:2020qpj} of the energy flux operators reveals that the collinear behavior of EEC is determined by the spin-3 non-local operators \cite{Hofman:2008ar}. Recently, the squeezed limit of EEEC has been investigated, where the light-ray OPE is developed at leading twist in QCD, in order to understand the transverse spin structure in the squeezed limit \cite{Chen:2021gdk}. In fact, this spin structure gives rise to a quantum interference at colliders: when rotating the squeezed detector by an angle $\phi$ with respect to the third detector, the interference between the intermediate virtual gluon with different helicity leads to a $\cos(2\phi)$ dependence \cite{Chen:2020adz}. Furthermore, standard CFT tools like conformal blocks and Lorentz inversion formula \cite{Caron-Huot:2017vep, Simmons-Duffin:2017nub} are also developed to organize the power correction of triple-collinear EEEC \cite{Chen:2022jhb, Chang:2022ryc}, opening a new window to studying jet substructure. More recent progress can be found in \cite{Chen:2022swd,Lee:2022ige}.

An outline of this paper is as follows. In Section~\ref{sec:calculation_setup}, we introduce the calculation method for three-point energy correlators at leading order. Briefly speaking, we directly integrate the tree-level matrix elements over the four-particle phase space and express the result in terms of transcendental polylogarithmic functions. With our parameterization, the non-analytic structure in the phase space factorizes and EEEC is reduced to a two-fold integral that can be calculated directly. We discuss the structure of the analytic expression and the numerical checks in Section~\ref{sec:results}. In Section~\ref{sec:analysis}, we extract the equilateral limit, the triple collinear limit and the squeezed limit contributions. The analytic formula for equilateral EEEC and its endpoint behaviors is given for all partonic channels. For the triple collinear limit, we also present a method that allows us to directly extract the subleading power corrections from expanding the EEEC integrand. We summarize in Section~\ref{sec:summary}.

\section{Calculation setup}\label{sec:calculation_setup}
%TKACHOV198165
The leading order EEEC arises from the tree-level process $\gamma^{*}\to \text{4 partons}$. Given the appearance of the non-standard measurement function in Eq.~\eqref{eq:eeecdef}, it is not easy to directly apply the modern loop techniques like Integration-by-parts (IBP) \cite{CHETYRKIN1981159} and differential equations \cite{KOTIKOV1991158,Gehrmann:2000zt}. While for the cases that only involving one non-standard cut propagator like $\delta(x_1 - (1- \cos \theta_{ij})/2 )$, a method was proposed in Refs.~\cite{Dixon:2018qgp,Luo:2019nig,Gao:2020vyx} to allow for a generalized IBP reduction in \texttt{LiteRed}~\cite{Lee:2012cn,Lee:2013mka} and \texttt{Fire}~\cite{Smirnov:2008iw,Smirnov:2014hma}. The appearance of three non-standard cut propagators in Eq.~\eqref{eq:eeecdef} makes the application of the method in Refs.~\cite{Dixon:2018qgp,Luo:2019nig,Gao:2020vyx} much less efficient. Instead of trying to improve the efficiency of the same method, we take the EEEC definition Eq.~\eqref{eq:eeecdef} and calculate the phase space integral directly. The main feature of our method is appropriate parameterizations of the four-particle phase space $\textit{dPS}_4$ and the kinematic space $\{x_1,x_2,x_3\}$, which makes the direct integration possible. Since we only care about EEEC at LO, it is safe to perform the computation in the $d=4$ dimension.   

%Instead of following the calculation method in EEC \cite{Dixon:2018qgp,Luo:2019nig} and applying the modern loop techniques like Integration-by-parts \cite{CHETYRKIN1981159,TKACHOV198165,Lee:2013mka,Lee:2012cn} and differential equations \cite{KOTIKOV1991158,Gehrmann:2000zt}, we take the EEEC definition Eq.~\eqref{eq:eeecdef} and calculate the phase space integral directly. Given the IR finiteness of EEEC at LO, it is safe to perform the computation in $d=4$ dimension. The main feature of our method is the appropriate parameterization of the four-particle phase space $\textit{dPS}_4$ and the kinematic space $\{x_1,x_2,x_3\}$, which makes the direct integration possible. 

\subsection{Amplitudes and topology identification}

\begin{figure}[ht]
\centering
\includegraphics[scale=1.3]{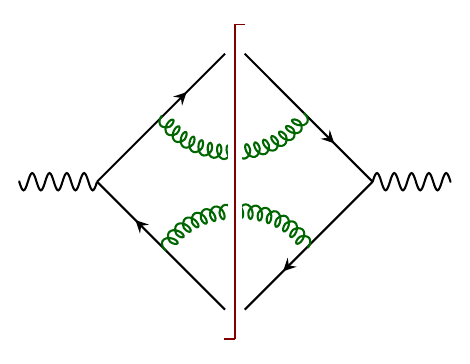}
\includegraphics[scale=1.3]{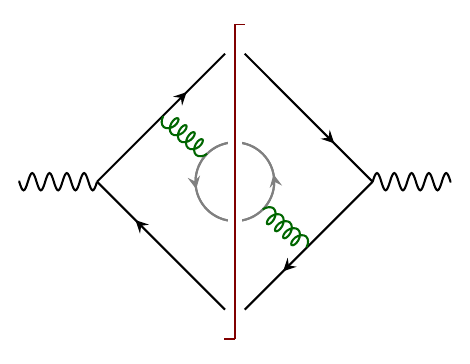}
\includegraphics[scale=1.3]{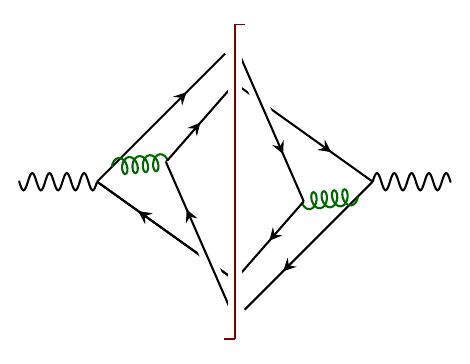}
\includegraphics[scale=1.3]{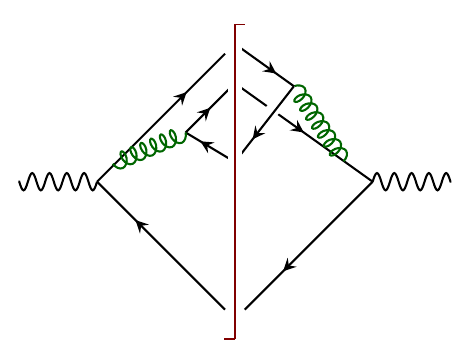}
\caption{Some typical graphs on the matrix elements squared $|\cM|^2$ for $\gamma^{*}\to \text{4 partons}$. The first graph corresponds to the double gluon emissions, while in the second graph, the gray lines represent the non-identical quark pair. The last two graphs show the interface between identical quark pairs.}
\label{fig:matrix_elements}
\end{figure}

We start by calculating the matrix elements squared $|\cM|^2$ for $\gamma^{*}\to \text{4 partons}$ with \texttt{QGRAF} \cite{NOGUEIRA1993279} and \texttt{FORM} \cite{Vermaseren:2000nd}, where the color algebra is handled by the \texttt{Color} package \cite{vanRitbergen:1998pn}. The calculation includes three subprocesses:
\begin{align}
\gamma^*(q)&\to q(p_1) +\bar q (p_2)+ q^\prime (p_3)+ \bar q^\prime (p_4)\,,\notag\\
\gamma^*(q)&\to q(p_1)+ \bar q (p_2) +q (p_3) +\bar q (p_4)\,,\notag\\
\gamma^*(q)&\to q(p_1) +\bar q (p_2) +g (p_3)+ g (p_4)\,,
\end{align}
where $q^\prime$ and $\bar q^\prime$ stand for non-identical quarks compared with the quarks $q$ and $\bar{q}$. In Fig.~\ref{fig:matrix_elements}, we present some typical diagrams for the matrix elements. We also compute the same matrix elements squared in \texttt{FeynArts} \cite{Hahn:2000kx} and \texttt{FeynCalc} \cite{Mertig:1990an,Shtabovenko:2016sxi} as a crosscheck. 
For both of them, we adopt the axial gauge when summing the gluon polarizations
\begin{equation}
\sum_{\lambda=1}^{2}\epsilon^\mu (p_i,\lambda)\epsilon^{* \nu}(p_i,\lambda)=-g^{\mu\nu}+\frac{\bar{n}^{\mu} p_i^{\nu} +\bar{n}^\nu p_i^\mu}{\bar{n} \cdot p_i}- \frac{\bar{n}^2p_i^{\mu} p_i^{\nu}}{(p_i\cdot \bar{n})^2}\,,
\end{equation}
where for a particular parton with the momentum $p_i$, the momentum of another parton $p_j$ is used as the auxiliary vector $\bar{n}$. By Lorentz invariance, the matrix elements squared are expressed in terms of the standard Mandelstam variables $s_{ij}=(p_i+p_j)^2$.

Our topology identification is a bit different from standard QCD calculations, where the established methods require the $\mathcal{UF}$-representation of the Feynman integrals \cite{Pak:2011xt}. In our calculation, it is enough to permute the final state momenta $p_{1,2,3,4}$ or equivalently, permute $s_{ij}$, and classify terms that are invariant under such transformations. Importantly, we have to carry the energy weights $E_i E_j E_k$ together since they are not invariant under particle renaming. Since we are not going to use the topologies for IBP reduction, the point of topology identification is to reduce the integrand and simplify the phase space integration. After obtaining the reduced matrix elements $|\cM(p_1,p_2,p_3,p_4)|^2$, we rename the particles such that the energy weights all become $E_1 E_2 E_3$:
\begin{align}
\label{eq:topoidexample}
    &\sum_{i\neq j\neq k \in \{ 1\,,2\,,3\,,4\} }\int \frac{E_iE_jE_k}{Q^3}{\it dPS}_4 \Pi_{ijk} |\cM(p_1,p_2,p_3,p_4)|^2\notag\\
    =&\sum_{a \neq b \neq c \in \{1\,,2\,,3\}}\int \frac{E_a E_b E_c}{Q^3} {\it dPS}_4 \Pi_{abc}\left(|\cM(p_a,p_b,p_c,p_4)|^2+|\cM(p_a,p_b,p_4,p_c)|^2\right.\notag\\
    &\hspace{4cm}\left.+|\cM(p_a,p_4,p_c,p_b)|^2+|\cM(p_4,p_a,p_b,p_c)|^2\right)\notag\\
    =&\bigg[\int \frac{E_1E_2E_3}{Q^3} {\it dPS}_4 \Pi_{123}\left(|\cM(p_1,p_2,p_3,p_4)|^2+|\cM(p_1,p_2,p_4,p_3)|^2+|\cM(p_1,p_4,p_3,p_2)|^2\right.\notag\\
    &\hspace{4cm}\left.+|\cM(p_4,p_1,p_2,p_3)|^2\right)\bigg]+\text{permutations of $x_1\,,x_2\,,x_3$}\,,
\end{align}
where $\Pi_{ijk}$ is a short-hand notation for the measurement function
\begin{equation}
\label{eq:measurementfunction}
    \Pi_{ijk}=\delta\left(x_1-\frac{1-\cos\theta_{jk}}{2}\right) \delta\left(x_2-\frac{1-\cos\theta_{ik}}{2}\right) \delta\left(x_3-\frac{1-\cos\theta_{ij}}{2}\right)\,.
\end{equation}
In the second line of Eq.~\eqref{eq:topoidexample}, we make the summation on $4$ explicit and rename the momentum labels such that there is no energy weight $E_4$ in the expression. Then we only need to calculate the unsymmetrical part in the last line since the result is fully symmetric in $x_1,x_2,x_3$ ($x_{1,2,3}$).

\subsection{Phase space parameterization}

The most challenging part of calculating EEEC is the calculation of the phase space integral. Recall that the massless four-particle phase space measure in $d$ dimension \cite{Gehrmann-DeRidder:2003pne} is given by
\begin{multline}
{\it dPS}_4= (2\pi)^{4-3d} (Q^2)^{3-\frac{d}{2}} 2^{1-2d} \delta(Q^2-s_{12}-s_{13}-s_{14}-s_{23}-s_{24}-s_{34})\\
\times (-\Delta_4)^{\frac{d-5}{2}}\Theta(-\Delta_4)  d\Omega_{d-1}d\Omega_{d-2}d\Omega_{d-3}ds_{12}ds_{13}ds_{14}ds_{23}ds_{24}ds_{34}\,,
\end{multline}
where the Gram determinant is
\begin{equation}
\Delta_4 = \lambda (s_{12}s_{34}, s_{13}s_{24}, s_{14}s_{23}),\quad \lambda(x, y, z)= x^2+y^2+z^2-2(x y +x z+y z)\,,
\end{equation}
and the $d$-dimensional hypersphere measure $d\Omega_d$ satisfies
\begin{equation}
V(d)=\int d\Omega_d = \frac{2\pi^{d/2}}{\Gamma(d/2)}\,.
\end{equation}

The main difficulty of the direct integration method then comes from the non-trivial constraint $\Theta(-\Delta_4)$, which corresponds to a complicated region of the four-particle phase space. To resolve this problem, we first introduce the energy fractions of three final state particles in the center of mass frame of $\gamma^*$,
\beq
z_1=\frac{2 p_1 \cdot q}{Q^2},\quad z_2= \frac{2 p_2 \cdot q}{Q^2}, \quad z_3= \frac{2 p_3 \cdot q}{Q^2}\,.
\eeq
Notice that $q = p_1 +p_2 +p_3+p_4$, the above equation becomes
\beq
z_1= s_{12}+s_{13}+s_{14},\quad z_2= s_{12}+s_{23}+s_{24}, \quad z_3=s_{13}+s_{23}+s_{34}\,.
\eeq
Together with EEEC measurement function in Eq.~\eqref{eq:measurementfunction}, all Mandelstam variables can be written in terms of three energy fractions and three kinematic variables,  
\begin{align}
&s_{12}=z_1 z_2 x_3,\quad s_{13}=z_1 z_3 x_2,\quad s_{23}=z_2 z_3 x_1,\nonumber \\
&s_{14}=z_1(1- z_2 x_3 - z_3 x_2),\quad s_{24}=z_2(1-z_1 x_3- z_3 x_1),\quad s_{34}=z_3(1- z_1 x_2 - z_2 x_1)\,,
\end{align}
where and in the following we set $Q^2 = 1$. Although the energy fractions break the symmetry of renaming final state particles, the complicated constraint $\Theta(-\Delta_4)$ decouples from the integrals. For example,
\begin{align}
\label{eq:egphasespace}
&\int ds_{12}ds_{13}ds_{14}ds_{23}ds_{24}ds_{34} \Theta(-\Delta_4) \delta(\sum_{i<j}s_{ij}-1)\Pi_{123} \nonumber \\
=&\int  dz_1 dz_2 dz_3 (z_1^2 z_2^2 z_3^2) \Theta(-\Delta_4) \frac{1}{1-x_2z_1-x_1z_2}\delta\left(z_3-\frac{z_1+z_2-x_3 z_1 z_2 -1}{z_1x_2+z_2x_1-1}\right) \nonumber \\
=&\Theta(-\widetilde{\Delta}_4) \int  dz_1 dz_2 dz_3 (z_1^2 z_2^2 z_3^2)\frac{1}{1-x_2z_1-x_1z_2}\delta\left(z_3-\frac{z_1+z_2-x_3 z_1 z_2 -1}{z_1x_2+z_2x_1-1}\right)\,, 
\end{align}
where $\Delta_4$ is factorized into integration variables dependent and non-dependent parts 
\begin{equation}
\frac{\Delta_4}{z_1^2 z_2^2 z_3^2} = x_1^2+x_2^2+x_3^2 - 2x_1x_2 -2x_1x_3 -2x_2x_3+4x_1x_2x_3 \equiv \widetilde{\Delta}_4 \,. 
\end{equation}
Here $\widetilde{\Delta}_4\leq 0$ becomes the constraint for the kinematic space $\{x_1,x_2,x_3\}$. Fig.~\ref{fig:kinematic_region} shows the allowed kinematic regions, and as we will see in Section~\ref{sec:analysis}, the shape dependence of EEEC is encoded in different limits of this region. Note that in the triple collinear limit, $\widetilde{\Delta}_4$ is further reduced to 
\begin{equation}\label{eq:collinearDelta4}
    \widetilde{\Delta}_4 \approx \widetilde{\Delta}_4^{\text{coll}}= x_1^2+x_2^2+x_3^2 - 2x_1x_2 -2x_1x_3 -2x_2x_3 \,,
\end{equation}
where $\sqrt{x_1}$, $\sqrt{x_2}$ and $\sqrt{x_3}$ can be interpreted as the lengths of three sides for a triangle due to Helen's area formula.

%\begin{figure}[ht]
%\centering
%\includegraphics[scale=0.6]{parameterspace.pdf}
%\caption{The ``zongzi''-shaped kinematic space $\{x_1,x_2,x_3\}$, which is constrained by the Heaviside function $\Theta(-\widetilde{\Delta}_4)$.}
%\label{fig:parameter}
%\end{figure}

In summary, EEEC is simplified to an integral over the energy fraction $z_1$, $z_2$ and $z_3$. While $z_3$ is integrated by the $\delta$ function in Eq.~\eqref{eq:egphasespace}, the remaining two-fold integral can be finished using a package called \texttt{HyperInt} \cite{Panzer:2014caa}.

%While $z_3$ can be integrated with the $\delta$ function in Eq.~\eqref{eq:egphasespace}, the $z_2$ integration can be easily finished with a standard mathematical tool. Then the computation of the remaining one-fold integral for $z_1$ requires us to use a more specific package like \texttt{HyperInt} \cite{Panzer:2014caa}.

\subsection{Direct integration}
Without the constraint from the $\delta$ function in Eq.~\eqref{eq:egphasespace},   
the ranges for integration variables $z_1$ and $z_2$ are both from 0 to 1. With the constraint, the integration regions become non-trivial. Explicitly, the result is found to be
\begin{align}
\label{eq:boundz1z2}
&\int  dz_1 dz_2 dz_3  \delta\left(z_3-\frac{z_1+z_2-x_3 z_1 z_2 -1}{z_1x_2+z_2x_1-1}\right) f(x_1,x_2,x_3,z_1,z_2,z_3) \nonumber \\
&= \int_0^1 dz_1 \int_0^{\frac{1-z_1}{1- x_3 z_1}} dz_2 f\left(x_1,x_2,x_3,z_1,z_2,\frac{z_1+z_2-x_3 z_1 z_2 -1}{z_1x_2+z_2x_1-1}\right)\,,
\end{align}
where we use $f(x_1,x_2,x_3,z_1,z_2,z_3)$ to represent the EEEC integrand. In our calculation, the integration of $z_2$ in Eq.~\eqref{eq:boundz1z2} can be easily carried out with standard mathematical tools like \texttt{Mathematica} or \texttt{Maple}. From the result, we identify the following two possible square roots that will appear in the final result of the LO EEEC,  
\begin{align}
\label{eq:sqrtEEEC}
&\sqrt{x_1^2+x_2^2+x_3^2-2x_1x_2-2x_1x_3-2x_2x_3} = \sqrt{ \widetilde{\Delta}_4^{\text{coll}}} \,, \nonumber \\
&\sqrt{x_1^2+x_2^2+x_3^2-2x_1x_2-2x_1x_3-2x_2x_3+4x_1x_2x_3}=\sqrt{\widetilde{\Delta}_4}\,. 
\end{align}
To perform the computation of the remaining one-fold integral with respect to $z_1$, we need to rationalize the square roots in Eq.~\eqref{eq:sqrtEEEC} by parameterizing the kinematic space $\{x_1,x_2,x_3\}$. Explicitly, we introduce a complex variable $z$ and its congugate $\bar{z}$ as well as a purely imaginary variable $t$ via
\begin{equation}\label{eq:zzbardef}
\frac{x_1}{x_3}=z \bar z,\quad \frac{x_2}{x_3}=(1-z)(1-\bar z),\quad x_3=\frac{t^2-(z-\bar z)^2}{4z \bar z (1-z)(1-\bar z)}\,,
\end{equation}
such that the two square roots are rationalized
\begin{align}
&\sqrt{x_1^2+x_2^2+x_3^2-2x_1x_2-2x_1x_3-2x_2x_3}=x_3 (z-\bar z)=\frac{t^2-(z-\bar z)^2}{4z \bar z (1-z)(1-\bar z)}  (z-\bar z) \,, \\
&\sqrt{x_1^2+x_2^2+x_3^2-2x_1x_2-2x_1x_3-2x_2x_3+4x_1x_2x_3}=x_3 t=\frac{t^2-(z-\bar z)^2}{4z \bar z (1-z)(1-\bar z)} t \,.
\end{align}
Note that as observed in Eq.~\eqref{eq:collinearDelta4} and in Ref.~\cite{Chen:2019bpb}, the second square root disappears in the triple collinear limit and we no longer need $t$ variable. While in the triple collinear limit, $z$ turns out to be a nice variable that characterizes the triangle shape dependence of EEEC and manifests the $\mathbb{S}_3\times \mathbb{Z}_2$ symmetry, $\{z,t\}$ are not good variables for the full shape dependence and for phenomenological studies eventually. So we will change back to the angular distances $x_{1,2,3}$ after finishing the calculation.

Using the ${z,t}$ parameterization, we partial fraction the integrand and format the denominators to be linear functions in the last integration variable $z_1$. Subsequently we can evaluate the final integration in \texttt{HyperInt} \cite{Panzer:2014caa}. It gives us the result in terms of Goncharov polylogarithms (GPLs) \cite{goncharov1mpl,Goncharov:1998kja,Borwein:1999js}, up to transcendentality-two. The GPL is defined iteratively by 
\begin{equation}\label{eq:gpl_def}
G(a_1,\cdots a_n; x)\equiv\int_0^x \frac{dt}{t-a_1} G(a_2,\cdots a_n; t)\,,
\end{equation}
with
\begin{equation}
G(;x)\equiv1,\quad G(\vec 0_n;x)\equiv\frac{1}{n!}\ln^n (x)\,.
\end{equation}
There are lots of analytic calculations at two-loop order found involving GPLs, both loop integrals and phase space integrals. 
%(e.g. \cite{Goncharov:2010jf,Duhr:2012fh,Anastasiou:2014vaa,Abreu:2019rpt,Lee:2021iid}). 
It is conjectured that GPLs up to transcendentality-three can be expressed in term of logarithms and classical polylogarithms $\Li_n(x)$ with $n\leq 3$ \cite{Duhr:2011zq}. For transcendentality-four GPLs, one also need the special function $\Li_{2,2}(x,y)$. For EEEC at LO, we only need GPLs up to transcendentality-two. The conversion from low transcendental weight GPLs to polylogarithms can be done with public packages like \texttt{PolyLogTools} \cite{Duhr:2019tlz} or \texttt{gtolrules.m} \cite{Frellesvig:2016ske}.
%or just perform the integral in the definition Eq.~\eqref{eq:gpl_def}.
We use the latter package to achieve the conversion. The same results can be obtained by the direct integration from the definition in Eq.~\eqref{eq:gpl_def}, and we also modify the arguments to meet \texttt{Mathematica}'s branch prescription for polylogarithms. After the conversion, our results are expressed in terms of classical polylogarithms.

To simplify the expression, we first collect the transcendental functions with the same rational coefficients. This constructs a raw transcendental function space in terms of classical polylogarithms. All the rational functions are simplified by the \texttt{MultivariateApart} package \cite{Heller:2021qkz}, which implements the partial fraction algorithms for multiple variables. However, simplifying the raw transcendental function space is in general not easy given the three variables. We start by applying transcendentality-two identities to simplify the individual base. A typical set of dilogarithm identities is as follows:
\begin{align}
\green{ \text{Reflection:}} \quad \Li_2(x)&=-\Li_2(1-x)-\log(x)\log(1-x)+\zeta_2\,, \notag\\
\green{ \text{Inversion:}}  \quad \Li_2(x)&=-\Li_2\left(\frac{1}{x}\right)-\frac{1}{2}\log^2(-x)-\zeta_2 \,, \notag\\
\green{ \text{Duplication:}} \quad  \Li_2(x)&=-\Li_2(-x)+\frac{1}{2}\Li_2(x^2)\,,
\end{align}
which all comes from the well-known five-term identity \cite{wojtkowiak1996functional}:
\begin{multline}
    \Li_2(x) + \Li_2(y) + \Li_2\left(\frac{1-x}{1-xy}\right) + \Li_2(1-xy) + \Li_2\left(\frac{1-y}{1-xy}\right) \\*
    = \frac{\pi^2}{2} - \log(x)\log(1-x)  - \log(y)\log(1-y) - \log\left(\frac{1-x}{1-xy}\right) \log\left(\frac{1-y}{1-xy}\right)\,.
\end{multline}
It turns out useful to use the five-term identity to simplify complicated arguments.
Then we add back all permutation terms of $x_{1,2,3}$ (what we call \textit{symmetrization}) as specified in Eq.~\eqref{eq:topoidexample}, and reorganize the result such that all bases and the corresponding coefficients are real. A better transcendental basis was already presented in Ref.~\cite{Yan:2022cye} for the $\cN=4$ SYM EEEC at LO. So for the last step, we try to project our function basis to the basis in Ref.~\cite{Yan:2022cye}. Explicitly, we symmetrize the $\cN=4$ function basis and construct a new linear independent transcendental weight-two basis. By evaluating both basis at a same numerical point and applying \texttt{PSLQ} algorithm \cite{PSLQref,Bailey:1999nv}, we managed to find the linear relations between the elements of these two function bases and successfully simplify our full result in QCD. As a crosscheck, we evaluate the original result from \texttt{HyperInt} numerically using public GPL libraries like \texttt{GiNac} \cite{Bauer:2000cp} and \texttt{FastGPL} \cite{Wang:2021imw}, and compare with the predictions of our final analytic expression.

It is interesting to ask how we can simplify the expression in the first step. On the one hand, this requires one to know the singularities of the result and to rule out all spurious poles and branch cuts. Landau equation \cite{LANDAU1959181} or Polynomial reduction \cite{arxiv.0910.0114} provides a sufficient set of possible singularities, but it is challenging to figure out the minimal set, especially in the high-dimensional complex hyperplane. Some of the progress can be found in Refs.~\cite{Bourjaily:2020wvq, Hannesdottir:2021kpd}. 
On the other hand, given the singularities of the integrals, there are still ambiguities how to choose the arguments of polylogarithms. To our knowledge, there is no public algorithm to search for the best arguments that make the expression shortest. It is possible that this can be done with \textit{symbol} \cite{Goncharov:2010jf} or even with \textit{Machine Learning} \cite{Dersy:2022bym} in the future, but it is out of scope of this paper.

\section{Results}\label{sec:results}

In this section, we present the full result for three-point energy correlator in QCD, in terms of the angular distance variables $x_{1,2,3}$ and the short-hand notations for the square roots:
\begin{align}
s_1& = \sqrt{\widetilde{\Delta}_4^{\text{coll}}}=\sqrt{x_1^2+x_2^2+x_3^2-2x_1x_2-2x_1x_3-2x_2x_3} \,,\\
s_2& =\sqrt{\widetilde{\Delta}_4} =\sqrt{x_1^2+x_2^2+x_3^2-2x_1x_2-2x_1x_3-2x_2x_3+4x_1x_2x_3}\,.
\end{align}
 At LO, there are three color channels
\begin{equation}
\frac{1}{\sigma_{\text{tot}}}\frac{d^3\sigma}{dx_1dx_2dx_3}=\left(\frac{\alpha_s}{4\pi}\right)^2 \frac{1}{4\pi \sqrt{-s_2^2}} \left(C_F T_F n_f H_{n_f}+C_F^2 H_{C_F}+C_F C_A H_{C_A}\right)\,,
\end{equation}
where we normalize the distribution to the born-level cross section, and $\alpha_s = g_s^2/(4 \pi)$ with $g_s$ being the strong coupling constant. All channels contain functions up to transcendentality-two and take the form
\begin{equation}
H\equiv H^{(0)} +H^{(1)} +H^{(2)} =H^{(0)}+\sum_{i=1}^{7}R_i^{(1)} f_i+\sum_{i=1}^{21}R_i^{(2)} g_i \,.
\end{equation}
The EEEC function space is composed of 7 logarithmic bases $f_{1,\cdots,7}$ and 21 polylogarithmic bases $g_{1,\cdots,21}$. The transcendental weight-one bases are 
\begin{align}
f_1&=\log (1-x_1),\quad f_2=\log x_1, \quad f_3=\log (1-x_2),\quad f_4=\log x_2, \quad f_5=\log (1-x_3),\notag\\
f_6&=\log x_3, \quad f_7=\log\left(2-s_2-x_1-x_2-x_3\right)-\log\left(2+s_2-x_1-x_2-x_3\right)
\end{align}
with the explicit $\mathbb{S}_3$ permutation symmetry. The transcendental weight-two bases are 
\begin{align}
g_1&=\text{Li}_2\left(\frac{x_1}{x_1-1}\right),\quad g_2=\text{Li}_2\left(\frac{x_2}{x_2-1}\right),\quad g_3=\text{Li}_2\left(\frac{x_3}{x_3-1}\right),\notag\\
g_4&=2 \re\left[
   \text{Li}_2\left(\frac{s_2+x_1-x_2+2 x_2 x_3-x_3}{2
   \left(x_2-1\right) \left(x_3-1\right)}\right)-\text{Li}_2\left(\frac{2 x_1 x_2}{s_2-x_1+2 x_1
   x_2-x_2+x_3}\right)\right]\notag\\
   &+2
   \text{Li}_2\left(\frac{x_1}{x_1-1}\right)-2
   \text{Li}_2\left(\frac{x_3}{x_3-1}\right),\notag\\
g_5&=g_4(x_2\leftrightarrow x_3),\notag\\
g_6&=-2 i \im\left[\text{Li}_2\left(\frac{2 \left(x_1-1\right)
   x_2}{s_2-x_1+2 x_1 x_2-x_2+x_3}\right)\right]\notag\\
   &\hspace{1.0cm}-2 i \im\left[\log
   \left(\frac{s_2-x_1+x_2-x_3}{2 x_1
   \left(x_2-1\right)}\right)\right] \re\left[\log \left(\frac{2
   \left(x_1-1\right) x_2}{s_2-x_1+2 x_1 x_2-x_2+x_3}\right)\right],\notag\\
g_7&=g_6(x_2\leftrightarrow x_3),\quad g_8=g_6(x_1,x_2,x_3\leftrightarrow x_2,x_3,x_1),\notag\\
g_9&=2 i \im\bigg[\text{Li}_2\left(\frac{s_2+x_1-x_2+
   x_3}{2(1- x_2)}\right)-\text{Li}_2\left(\frac{s_2+x_1-x_2-x_3}{2
   \left(x_1-1\right)}\right)-\text{Li}_2\left(\frac{2 x_1
   x_2}{s_2+x_1+x_2-x_3}\right)\notag\\
   &+\frac{1}{2} \log \left[\left(1-x_1\right)
   \left(x_2-1\right) \left(x_3-1\right)\right] \log
   \left(2-s_2-x_1-x_2-x_3\right)\bigg],\notag\\
g_{10}&=\pi^2,\quad g_{11}=-4\left[ \im\left[\log \left(2-s_2-x_1-x_2-x_3\right)\right]\right]^2,\notag\\
g_{12}&=2 i \log \left(\frac{x_2\left(x_1-1\right)}{x_1
   \left(x_2-1\right)}\right) \im\left[\log
   \left(-s_2-x_1-x_2-x_3+2\right)\right],\quad g_{13}=g_{12}(x_2\leftrightarrow x_3),\notag\\
g_{14}&=\log \left(1-x_1\right) \log \left[\frac{\left(x_1-1\right)
   x_2}{x_1 \left(x_2-1\right)}\right]-\log \left(1-x_3\right) \log
   \left[\frac{x_2 \left(x_3-1\right)}{\left(x_2-1\right)
   x_3}\right],\notag\\
g_{15}&=g_{14}(x_2\leftrightarrow x_3),\quad g_{16}=g_{14}(x_1\leftrightarrow x_2),\notag\\
g_{17}&=-i
   \im\left[\text{Li}_2\left(\frac{s_2+x_1+x_2+x_3-2}{-s_2+x_1+x_2+
   x_3-2}\right)\right]\notag\\
   &\hspace{1cm}-i \log
   \left[\frac{s_2^2}{\left(x_1-1\right) \left(x_2-1\right)
   \left(x_3-1\right)}\right] \im\left[\log
   \left(2-s_2-x_1-x_2-x_3\right)\right],\notag\\
g_{18}&=\text{Li}_2\left(\frac{x_1-x_2}{x_1(1- x_2)}\right)+\frac{1}{2}
   \log \left[\frac{x_2 \left(x_1-1\right) }{x_1
   \left(x_2-1\right)}\right] \log \left[\frac{x_3}{x_1(1-x_2)}\right],\notag\\
g_{19}&=g_{18}(x_2\leftrightarrow x_3), \quad g_{20}=g_{18}(x_1,x_2,x_3\leftrightarrow x_2,x_3,x_1),\notag\\
g_{21}&=-2 i
   \bigg\{\im\bigg[\text{Li}_2\left(\frac{s_1+x_1-x_2-x_3}{s_2+x_1-x_2-x_3}\right)-\text{Li}_2\left(\frac{s_1+x_1+x_2-x_3}{-s_2+x_1+x_2-x_3}\right)\notag\\
   &-\text{Li}_2\left(\frac{s_1+x_1+x_2-x_3}{s_2+x_1+x_2-x_
   3}\right)-\text{Li}_2\left(-\frac{s_1-x_1+x_2+x_3}{s_2+x_1-x_2-x_3}\right)\notag\\
   &-\text{Li}_2\left(\frac{2
   \left(x_1-1\right) x_2 \left(s_1+x_1-x_2+x_3\right)}{\left(s_2+x_1-2 x_1 x_2+x_2-x_3\right)
   \left(s_2-x_1+x_2+x_3\right)}\right)\notag\\
   &-\text{Li}_2\left(\frac{2 \left(x_1-1\right) x_2
   \left(s_1+x_1-x_2+x_3\right)}{\left(s_2+x_1-x_2-x_3\right) \left(s_2-x_1+2 x_1
   x_2-x_2+x_3\right)}\right)\bigg]\notag\\
   &+\im\left[\log \left(2-s_2-x_1-x_2-x_3\right)\right]\log
   \left(\frac{s_1-s_2}{s_1+s_2}\right)\notag\\
   & -\im\left[\log
   \left(\frac{s_2-s_1}{s_2-x_1+x_2+x_3}\right)\right] \re\left[\log
   \left(\frac{s_1+x_1-x_2-x_3}{s_2+x_1-x_2-x_3}\right)\right]\notag\\
   &-\im\left[\log
   \left(\frac{s_1+s_2}{s_2-x_1-x_2+x_3}\right)\right] \re\left[\log
   \left(\frac{s_1+x_1+x_2-x_3}{-s_2+x_1+x_2-x_3}\right)\right]\notag\\
   &-\im\left[\log
   \left(\frac{s_2-s_1}{s_2+x_1+x_2-x_3}\right)\right] \re\left[\log
   \left(\frac{s_1+x_1+x_2-x_3}{s_2+x_1+x_2-x_3}\right)\right]\notag\\
   &-\im\left[\log
   \left(\frac{s_1+s_2}{s_2+x_1-x_2-x_3}\right)\right] \re\left[\log
   \left(\frac{-s_1-x_1+x_2+x_3}{s_2-x_1+x_2+x_3}\right)\right]\notag\\
   &+\im\left[\log \left(\frac{2
   \left(s_1-s_2\right) \left(x_1-1\right) x_2}{\left(s_2+x_1-2 x_1 x_2+x_2-x_3\right)
   \left(s_2-x_1+x_2+x_3\right)}\right)\right]\notag\\
   &\times\re\left[\log \left(\frac{2 \left(x_1-1\right) x_2
   \left(-s_1+x_1-x_2+x_3\right)}{\left(s_2+x_1-2 x_1 x_2+x_2-x_3\right)
   \left(s_2-x_1+x_2+x_3\right)}\right)\right]\notag\\
   &+\im\left[\log \left(\frac{2 \left(s_1+s_2\right)
   \left(x_1-1\right) x_2}{\left(s_2+x_1-x_2-x_3\right) \left(s_2-x_1+2 x_1
   x_2-x_2+x_3\right)}\right)\right]\notag\\
   &\times \re\left[\log \left(\frac{2 \left(x_1-1\right) x_2
   \left(s_1+x_1-x_2+x_3\right)}{\left(s_2+x_1-2 x_1 x_2+x_2-x_3\right)
   \left(s_2-x_1+x_2+x_3\right)}\right)\right]\bigg\}\,.
\end{align}
Although we use the short-hand notation $\re$ and $\im$ to make the expressions compact, all the bases are analytic functions themselves. 
%For example, 
%\begin{multline}
%2\re \left[\Li_2\left(\frac{s_2+x_1-x_2-x_3+2x_2x_3}{2(1-x_2)(1-x_3)}\right)\right]\\=\Li_2\left(\frac{s_2+x_1-x_2-x_3+2x_2x_3}{2(1-x_2)(1-x_3)}\right)+\Li_2\left(\frac{-s_2+x_1-x_2-x_3+2x_2x_3}{2(1-x_2)(1-x_3)}\right)
%\end{multline}
%is well-defined away from the branch cuts of the $\Li_2$ functions. 
The corresponding coefficients $R_i^{(1)}$ and $R_i^{(2)}$ are rational functions in terms of $x_{1,2,3}$ and $s_{1},s_2$. We provide these coefficients in the ancillary file.

We emphasize that EEEC encodes both scaling information and non-trival shape dependence since it is a three-parameter jet observable. Unlike collinear EEEC, the longest angular distance $x_L=\text{max}\{ x_1,x_2,x_3\}$ does not factorize out. Instead, the kinematic space is fully determined by the second square root $s_2^2\leq 0$, i.e. 
\begin{equation}\label{eq:kinematic_space}
x_1^2+x_2^2+x_3^2-2x_1x_2-2x_1x_3-2x_2x_3+4x_1x_2x_3<0 \,. 
\end{equation}

\begin{figure}[htb]
    \centering
    \includegraphics[scale=0.75]{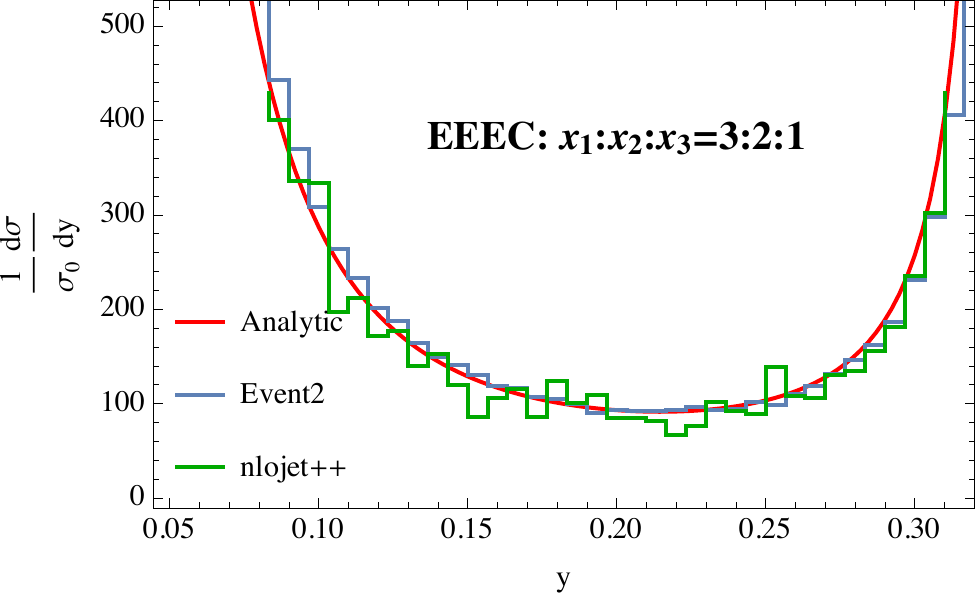}
    \includegraphics[scale=0.75]{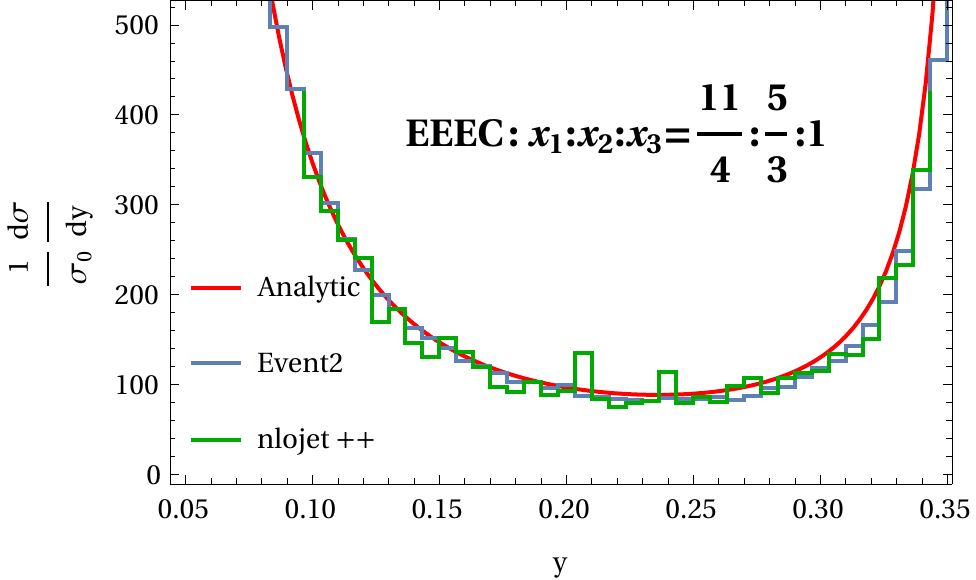}
    \caption{Comparison of the analytic result with the numerical programs \texttt{Event2} and \texttt{NLOJet++} for $\{x_1,x_2,x_3\}=\{3y,2y,y\}$ and $\{x_1,x_2,x_3\}=\{\frac{11}{4}y,\frac{5}{3}y,y\}$. Due to Eq.~\eqref{eq:kinematic_space}, the kinematic spaces are cutoff at $y= \frac{1}{3}$ and $y = \frac{959}{2640} \approx 0.36 $ respectively. Fifty billion points are sampled and the internal cutoff is set to $10^{-14}$ in \texttt{Event2}. Ten billion events are generated in \texttt{NLOJet++}.}
    \label{fig:eeec_full_check}
\end{figure}

\begin{figure}[htb]
    \centering
    \includegraphics[scale=0.55]{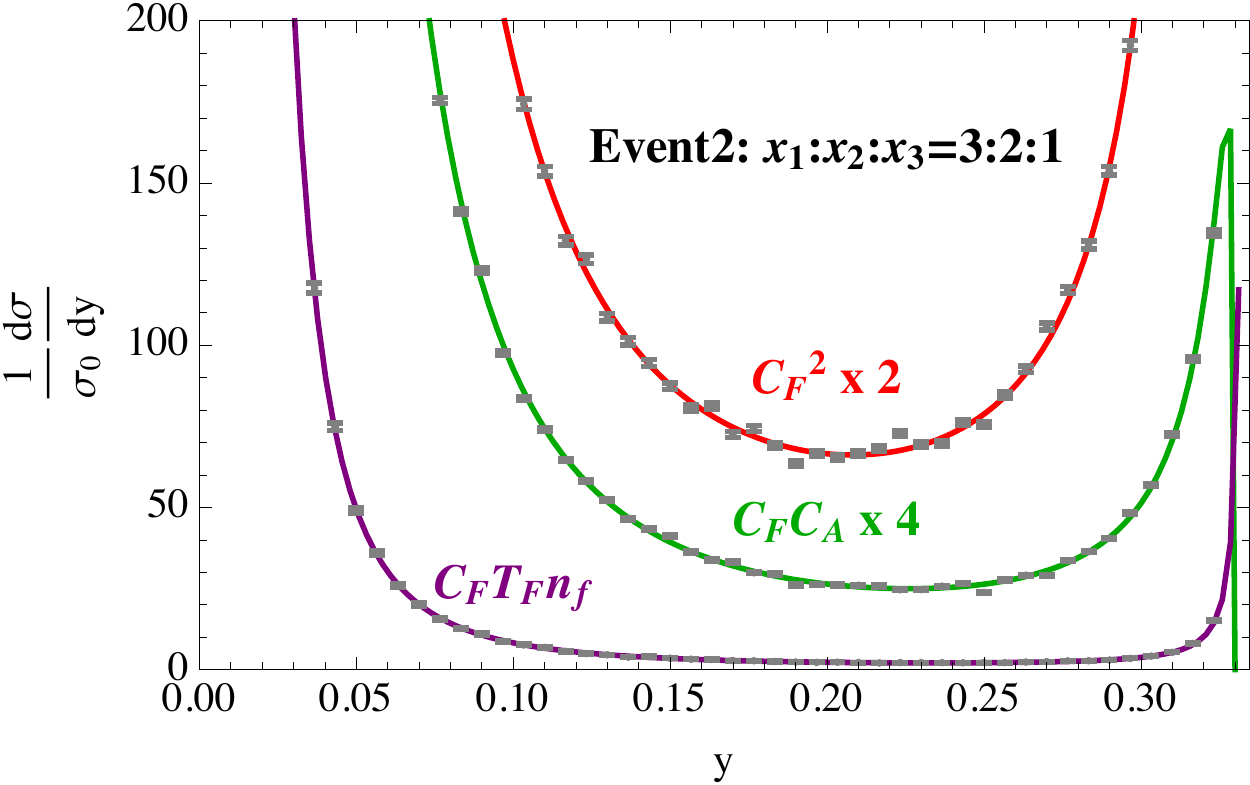}
    \includegraphics[scale=0.55]{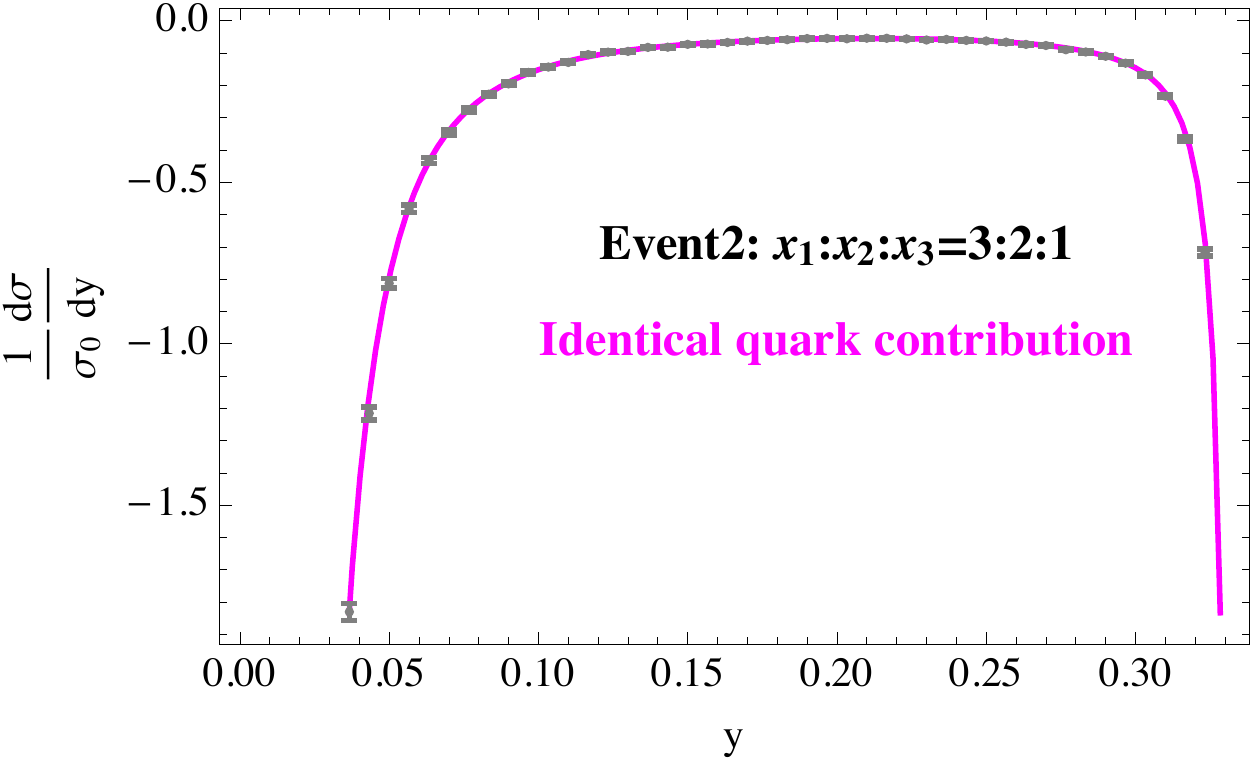}
    \caption{Left: Comparison of the analytic result with \texttt{Event2} for separate color structures with $\{x_1,x_2,x_3\}=\{3y,2y,y\}$. The $C_F^2$ and $C_F C_A$ are multiplied by a constant for clarity. Right: The identical quark part is verified separately since its contribution is small.}
    \label{fig:eeec_color_check}
\end{figure}

To verify our analytic result, we consider two special cases $\{x_1,x_2,x_3\}=\{3y,2y,y\}$ as well as $\{x_1,x_2,x_3\}=\{\frac{11}{4}y,\frac{5}{3}y,y\}$ and calculate them in \texttt{Event2} \cite{Catani:1996jh,Catani:1996vz} and \texttt{NLOJet++} \cite{Nagy:2003tz}. 
%To apply the constraint numerically, we introduce an error $\epsilon=10^{-3}$, such that
%\begin{equation}
%|x_1-x_2|<\epsilon,\quad |x_1-x_3|<\epsilon,\quad |x_2-x_3|<\epsilon
%\end{equation}
The obtained results are in good agreement with our analytic results (see Fig.~\ref{fig:eeec_full_check}). We also pick the first configuration and separate different color structures as well as the identical quark pair contribution in \texttt{Event2}. The detailed comparison can be found in Fig.~\ref{fig:eeec_color_check}.

Our result can be useful for phenomenological studies in precision QCD and jet physics. With a simple function basis as well as the simplified rational coefficients, evaluating its numerical values to high precisions is much faster than the raw GPL expression or a Monte Carlo program. As an example, it is easy to numerically evaluate our analytic expression in \texttt{Mathematica} to 200 digits precision within 4 seconds for a regular point in a single core machine. The simplicity of the result strongly encourages us to compute EEEC in QCD for gluon-initiated or $b\bar{b}$-initiated Higgs decays analytically in the future.

\section{Kinematic analysis}\label{sec:analysis}

\begin{figure}[htb]
    \centering
    \includegraphics[scale=1.0]{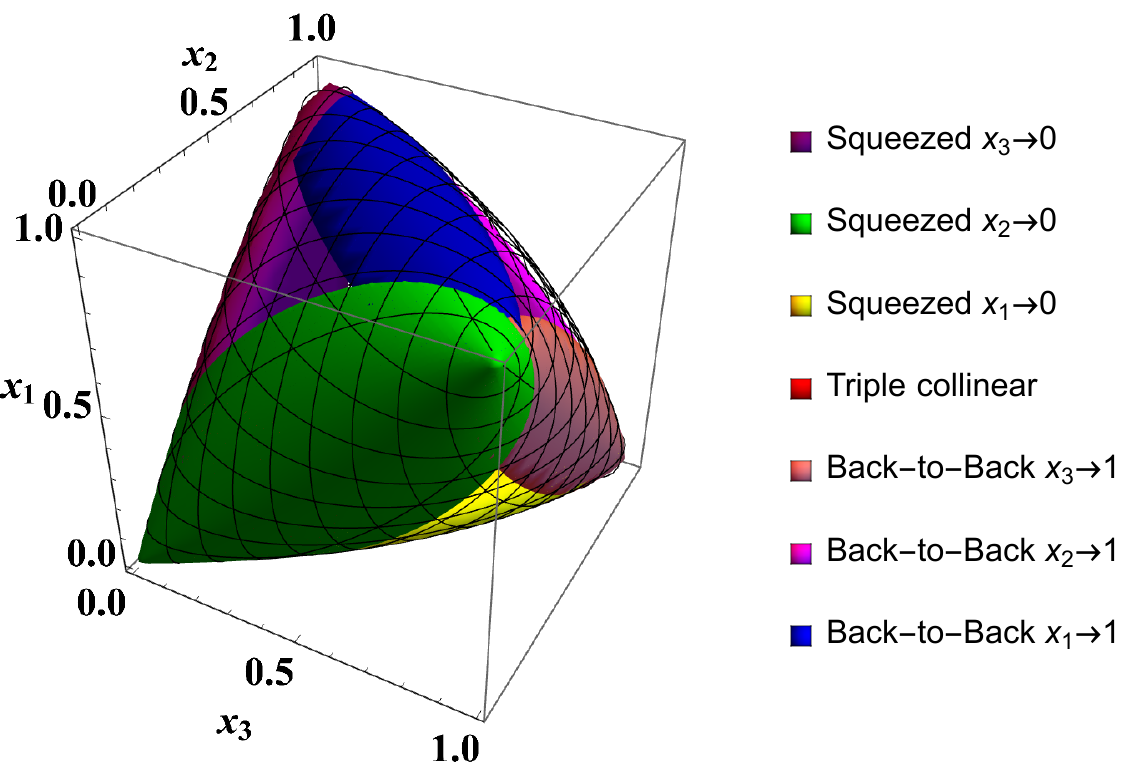}
    \caption{Various kinematic limits in the $\{x_1,x_2,x_3\}$ ``zongzi''-shaped space. We denote the triple-collinear limit, squeezed limits and back-to-back limits using different colors. The coplanar limit corresponds to the boundary of the kinematic space itself. The full 3D dynamic figure can be found in the ancillary file.}
    \label{fig:kinematic_x123}
\end{figure}

\begin{figure}[!htp]
	\centering
	\begin{subfigure}{.4\linewidth}
		\includegraphics[scale=0.3]{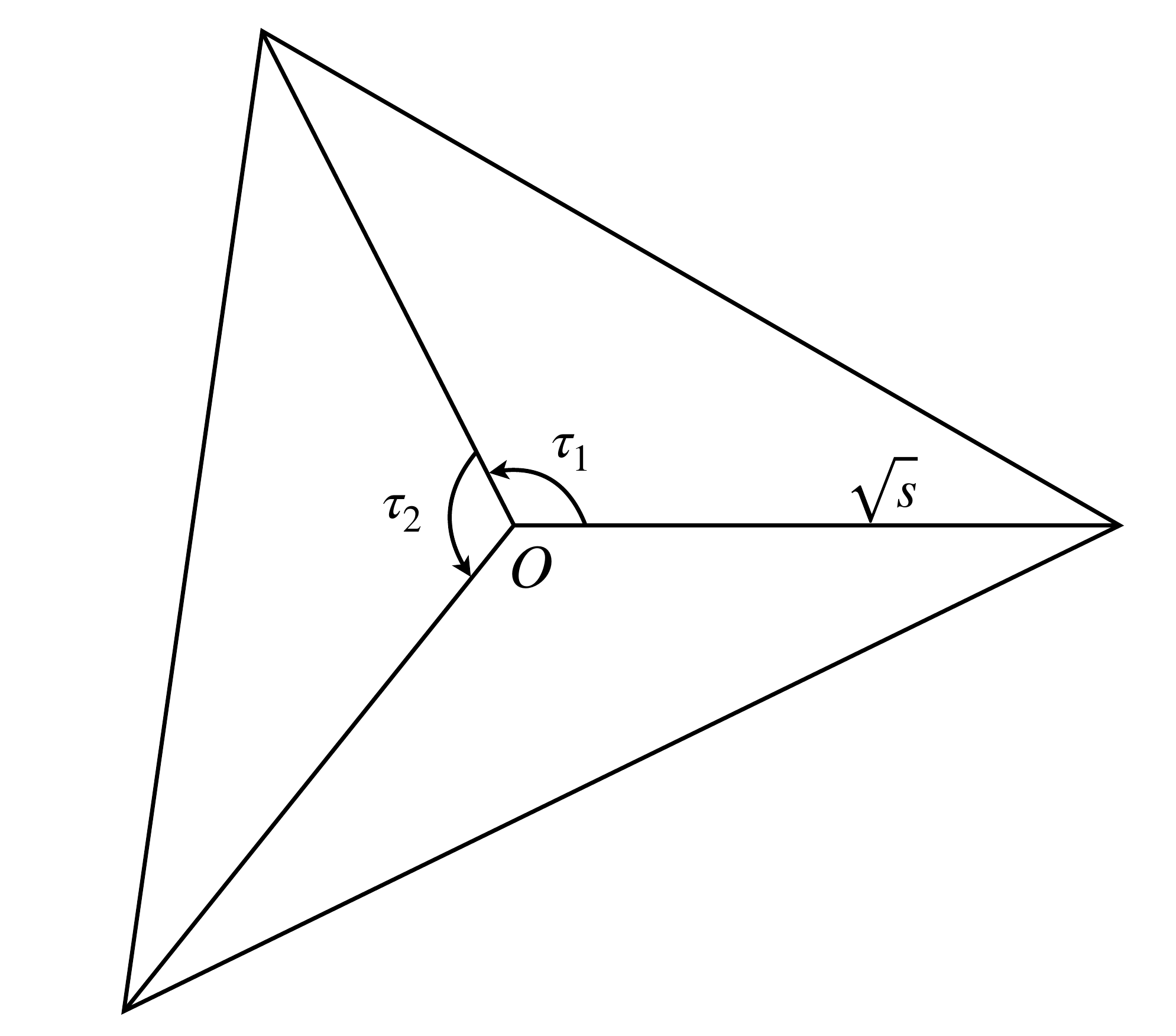}
		\label{fig:vars_plot}
		\caption{}
	\end{subfigure}
	\qquad\quad
	\begin{subfigure}{.4\linewidth}
		\includegraphics[scale=0.5]{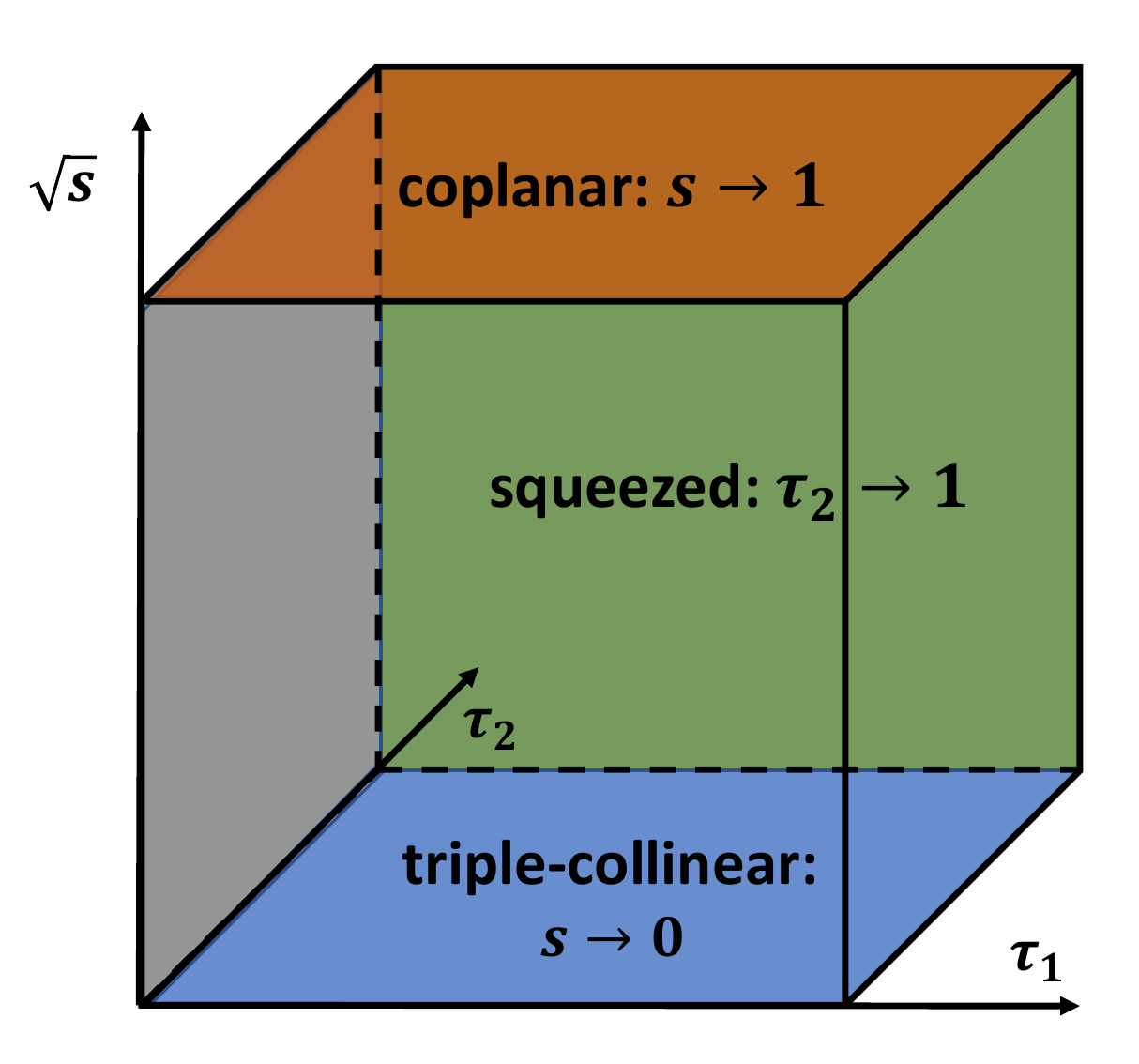}
		\label{fig:kinematic_region_vars}
		\caption{}
	\end{subfigure}
	\caption{(a) A graph on the $\{s,\tau_1,\tau_2\}$ coordinate on the celestial sphere \cite{Yan:2022cye}. (b) The kinematic limits of EEEC under the $\{s,\tau_1,\tau_2\}$ coordinate. $s\to 0$ and $s\to 1$ lead to the triple collinear and coplanar limit respectively, while $\tau_i\to1$ represents squeezed limit.}
	\label{fig:kinematic_stau}
\end{figure}

Given the complete shape dependence of the three-point energy correlator, it is interesting to investigate its behavior under different kinematic limits. Fig.~\ref{fig:kinematic_x123} shows several typical regions that could be useful for understanding the singularities and resummation. They are
\begin{itemize}
\item \green{Triple collinear limit:} $x_1\sim 0$, $x_2\sim 0$ and $x_3\sim 0$
\item \green{Squeezed limit:} $x_1\sim 0$, $x_2,x_3\sim x$ and its permutations
\item \green{Back-to-back limit:} $x_1\sim 1$ and its permutations
\item \green{Coplanar limit:} $s_2\to 0$
\end{itemize}
Alternatively, one can borrow the variables $\{s,\tau_1,\tau_2\}$ from \cite{Yan:2022cye}, which is related to the angular distance via 
\begin{equation}
x_1=-\frac{s}{(s+1)^2}\frac{(1-\tau_1)^2}{\tau_1},\quad x_2=-\frac{s}{(s+1)^2}\frac{(1-\tau_2)^2}{\tau_2},\quad x_3=-\frac{s}{(s+1)^2}\frac{(1-\tau_1 \tau_2)^2}{\tau_1 \tau_2}
\end{equation}
Here we put the three points in a circle with radius $\sqrt{s}$ on the celestial sphere and $\tau_{1,2}$ corresponds to the angle between two of them (see Fig.~\ref{fig:kinematic_stau}). One can also extract the kinematic limits using the new coordinate, and particularly, it is more convenient to expand the coplanar limit via $s\to 1$.

From phenomenological perspective, one can also slice the kinematic space and apply a constraint on the angular distance $x_{1,2,3}$. Since the demonstration of the full EEEC requires a 3D density plot, from which it is difficult to read information, applying the kinematic constraints helps to reduce the dimension of the plots. The most simplest case is the equilateral EEEC, where three angular distances are the same $x_1=x_2=x_3=x$. 
%In Fig.~\ref{fig:kinematic_x123_shape}, we also present the isosceles configuration $x_1=x_2$ and the right configuration $x_3+x_2=x_1$. 
Two other typical choices are isosceles configuration $x_1=x_2$ and the right configuration $x_3+x_2=x_1$.
When dealing with data from experiments or simulation programs like \texttt{Pythia} \cite{Sjostrand:2006za,Sjostrand:2014zea, Bierlich:2022pfr}, it is straightforward to apply these constraints directly in the event selection. There are also overlaps between the expansion of kinematic limits and the configuration constraints. For example, one can study both the triple collinear limit and coplanar limit in the equilateral configuration. Applying both expansion and slicing together gives a clearer picture on specific jet substructure that we want to understand.

%\begin{figure}[htb]
%    \centering
%    \includegraphics[scale=0.75]{shape_plot.png}
%    \caption{Examples of configuration constraints on EEEC. The blue solid line shows the equilateral EEEC, which only depends on a single variable. The purple and green regions represent the isosceles and right configurations respectively. The black solid line demonstrates their overlap.}
%    \label{fig:kinematic_x123_shape}
%\end{figure}

In this section, we will focus on the analytic result of equilateral EEEC and triple collinear EEEC at next-to-leading power, as well as the squeezed limit. We present the full expression for the equilateral EEEC, with equilateral function space included. To give a concrete example of applying both the kinematic expansion and configuration constraint together, we discuss the $x\to 0$ (collinear limit) and $x\to \frac{3}{4}$ (coplanar limit) singular behaviors, which could be interesting to the studies of trijet events at colliders. Regarding the triple collinear limit, we present the NLP correction analytically. It turns out that the collinear function space only contains one transcendental weight-one function and three weight-two functions under the $\mathbb{S}_3$ symmetry. Finally, we extract the LP squeezed limit and discuss the ambiguity of the definition under the triple collinear limit. The geometry reveals that the squeezed limit is actually path-dependent.

\subsection{Equilateral limit}

It is straightforward to extract the equilateral EEEC from our analytic formula. Alternatively, one can take the equilateral limit before performing the integral. Explicitly, Eq.~\eqref{eq:egphasespace} becomes
%\frac{1}{\sigma_{\text{tot}}}\frac{d^3\sigma}{dx^3}\sim
\begin{equation}
\Theta(3-4x) \int_0^1 dz_1 \int_0^{\frac{1-z_1}{1-x z_1}} dz_2 dz_3\frac{z_1^2 z_2^2 z_3^2}{1-z_1 x-z_2 x}\delta\left(z_3-\frac{z_1+z_2-x z_1z_2-1}{x z_1+x z_2 -1}\right)
\end{equation}
which can be evaluated directly. The Heaviside function suggests the equilateral EEEC is cutoff at $x=\frac{3}{4}$.

The analytic result is written as 
\begin{equation}
\label{eq:equilateral_limit}
\frac{1}{\sigma_0}\frac{d^3 \sigma}{dx^3}=\bigg(\frac{\alpha_s}{4\pi}\bigg)^2\mathcal{N}\times \bigg(G_{q \bar{q}q\prime \bar{q}\prime}(t) + \frac{1}{4}G_{q \bar{q}q \bar{q}}(t) + \frac{1}{2}G_{q \bar{q} g g}(t)\bigg)\,,
\end{equation}
where the normalization factor is $\mathcal{N}=\frac{1}{4\pi x \sqrt{3-4x}}$, and $\frac{1}{4}$ and $\frac{1}{2}$ are symmetry factors due to identical particles. In Eq.~\eqref{eq:equilateral_limit}, we also introduce another variable $t=\sqrt{3-4 x }$ to make the result more compact. The $n_f$ contribution is given by
\begin{align}
G_{q \bar{q}q\prime \bar{q}\prime}(t)&=C_f T_f n_f \bigg\{  -\frac{8}{2835 \left(t^2-3\right)^6 \left(t^2-1\right)^3}  \bigg( 872935 t^{16}+7645260 t^{14}-78741432 t^{12}\notag\\
&+174628460 t^{10}-57642594 t^8-167457660 t^6+106781904 t^4+50927940 t^2\notag\\
&-34837533  \bigg) +\frac{8}{8505 (t^2-1)^4 t \left(t^2-3\right)^7} \bigg(5275445 t^{22}+31825710 t^{20}-554071427 t^{18}\notag\\
&+1961298184 t^{16}-1956329238 t^{14}-2450875468 t^{12} +6441472482 t^{10}-3202455096 t^8\notag\\
&-2041671807 t^6+2142124110 t^4-315629055 t^2-34836480\bigg)\bigg(\pi -3 \tan ^{-1}(t)\bigg)\notag\\
&+\frac{4}{35 (t^2-1)^4 \left(t^2-3\right)^7} \bigg(43575 t^{20}+312270 t^{18}-3991645 t^{16}+10221960 t^{14}\notag\\
&-1074402 t^{12}-29837836 t^{10}+41621582 t^8-12477368 t^6 -14178453 t^4+12394094 t^2\notag\\
&-3141297\bigg) \log \left(\frac{t^2+1}{4} \right) +\mathcal{T}_2(t) \bigg\}\,,
\end{align}
where the transcendentality-two part $\mathcal{T}_2(t)$ is 
\begin{align}
\mathcal{T}_2(t)&=\frac{64 \left(11 t^4-774 t^2+135\right) }{81 \sqrt{3} \left(t^2-3\right)^2}g_1^{(2)} -\frac{3}{\left(t^2-3\right)^5 \left(t^2-1\right)^5}  \bigg(-415 t^{20}-4634 t^{18}+20033 t^{16}\notag\\
&-10488 t^{14}-51326 t^{12}+84452 t^{10}-19254 t^8-54136 t^6+48381 t^4-14682 t^2+3093\bigg)g_2^{(2)}\notag\\
&-\frac{3(t^2+1)}{\left(t^2-3\right)^5 \left(t^2-1\right)^5} \bigg(1657 t^{18}+16957 t^{16}-97956 t^{14}+145260 t^{12}+40262 t^{10} -334978 t^8\notag\\
&+365500 t^6-150900 t^4+16425 t^2-5811\bigg) g_3^{(2)}+\frac{24}{\left(t^2-3\right)^5 \left(t^2-1\right)^5} \bigg(207 t^{20}+2330 t^{18}\notag\\
&-10161 t^{16}+6136 t^{14}+22366 t^{12}-35044 t^{10}+1878 t^8+26744 t^6-14349 t^4 \nonumber \\
&-678 t^2-453\bigg)g_4^{(2)}\,,
\end{align}
with the corresponding function space
\begin{align}
g_1^{(2)}&=D_2^{-}\left(\frac{t-\sqrt{3}}{t-i}\right)-D_2^{-}\left(\frac{t+\sqrt{3}}{t-i}\right)+\frac{1}{3} \log \left(\frac{\sqrt{3}+t}{\sqrt{3}-t}\right) \left( \pi -3 \tan ^{-1}(t)\right)\,,\notag\\
g_2^{(2)}&=\text{Li}_2\left(\frac{t-\sqrt{3}}{-i+t}\right)+\text{Li}_2\left(\frac{t+\sqrt{3}}{-i+t}\right)+\text{Li}_2\left(\frac{t-\sqrt{3}}{i+t}\right)+\text{Li}_2\left(\frac{t+\sqrt{3}}{i+t}\right) \,, \notag\\
g_3^{(2)}&=2 (\tan ^{-1}t)^2+\zeta_2 \,, \notag\\
g_4^{(2)}&=\pi \tan ^{-1}(t) \,. 
\label{eq:basis1}
\end{align}
Here $D_2^{-}(z)$ is the \textit{Bloch-Wigner} function
\begin{equation}
2 i D_2^{-}(z)=\text{Li}_2(z)-\text{Li}_2\left(\bar z\right)+\frac{1}{2}
   \left(\log (1-z)-\log \left(1-\bar z\right)\right) \log \left(z \bar z\right)\,.
   \label{eq:blockwigner}
\end{equation}
As a Single-valued Harmonic Polylogarithm (SVHPL), Bloch-Wigner function satisfies 
\begin{equation}
D_2^-(z)=D^-_2\left(1-\frac{1}{z}\right)=D_2^-\left(\frac{1}{1-z}\right)=-D_2^-\left(\frac{1}{z}\right)=-D_2^-(1-z)=-D_2^-\left(\frac{-z}{1-z}\right)
\end{equation}
 and is parity-odd under $\mathbb{Z}_2$ symmetry. 

The results for the other two partonic channels are given as follows,
\begin{align}
G_{q \bar{q}q\prime \bar{q}\prime}(t)&=C_F(C_A-2C_F)\bigg\{\frac{32}{2835 \left(t^2-3\right)^6 \left(t^2-1\right)} \bigg(75565 t^{12}+3240230 t^{10}-17398269 t^8\notag\\
&+18684804 t^6+22325715 t^4-11993994 t^2-14389731\bigg)-\frac{32 t}{8505 \left(t^2-3\right)^7 \left(t^2-1\right)^2}\notag\\
& \bigg(329735 t^{16}+20536600 t^{14}-165711644 t^{12}+358668392 t^{10}-124678326 t^8\notag\\
&-343694808 t^6+231523236 t^4+95157720 t^2-65599065\bigg) \bigg(\pi -3 \tan ^{-1}(t)\bigg)\notag\\
&-\frac{16}{35 \left(t^2-3\right)^7 \left(t^2-1\right)^2}\bigg(2205 t^{16}+181860 t^{14}-1202880 t^{12}+1813700 t^{10}+303506 t^8\notag\\
&-676788 t^6-1242776 t^4+240652 t^2+553641\bigg) \log \bigg(\frac{t^2+1}{4}\bigg)+\mathcal{T}^{(id)}_2(t)\bigg\}\,,
\end{align}
\begin{align}
\mathcal{T}^{(id)}_2(t)&=\frac{12288t \left(3 t^8+2 t^6+116 t^4-66 t^2-55\right)}{\left(t^2-3\right)^7} g_7^{(2)}\notag\\
&+\frac{12288   t \left(t^8-2 t^6+36 t^4-30 t^2-5\right)}{\left(t^2-3\right)^7}g_8^{(2)}+\frac{12}{\left(t^2-3\right)^7 \left(t^2-1\right)^3}\bigg(99 t^{20}\notag\\
&+6454 t^{18}-56825 t^{16}+149896 t^{14}-72458 t^{12}-263676 t^{10}+401078 t^8-170360 t^6\notag\\
&+7847 t^4+11958 t^2-21181\bigg)g_9^{(2)}-\frac{96}{\left(t^2-3\right)^7 \left(t^2-1\right)^3}\bigg(13 t^{20}+794 t^{18}-6983 t^{16}\notag\\
&+18040 t^{14}-6294 t^{12}-40676 t^{10}+65098 t^8-40456 t^6+15625 t^4-3750 t^2-2435\bigg)g_4^{(2)}\notag\\
&+\frac{24}{\left(t^2-3\right)^7 \left(t^2-1\right)^3}\bigg(99 t^{20}+6454 t^{18}-56825 t^{16}+149896 t^{14}-72458 t^{12}-259580 t^{10}\notag\\
&+388790 t^8-162168 t^6+16039 t^4-330 t^2-17085\bigg)(\tan ^{-1}t)^2-\frac{128}{81 \sqrt{3} \left(t^2-3\right)^7}\notag\\
& \bigg(7 t^{14}-195 t^{12}+4563 t^{10}+53649 t^8+251829 t^6+337527 t^4-541647 t^2-102789 \bigg)g_1^{(2)}\notag\\
&-\frac{12 \left(21 t^{16}+1816 t^{14}-4220 t^{12}-8 t^{10}+3238 t^8+3976 t^6-7084 t^4+2280 t^2-275\right)}{\left(t^2-3\right)^5 \left(t^2-1\right)^3}g_2^{(2)}\notag\\
&-\frac{36864 t \left(t^2-1\right) \left(t^6-t^4+27 t^2-3\right) }{\left(t^2-3\right)^7}g_5^{(2)}
+\frac{589824 t \left(t^2-1\right) \left(t^2+1\right)}{\left(t^2-3\right)^7}g_6^{(2)} \,,
\end{align}
\begin{align}
G_{q \bar{q} g g}(x)&=C_F^2\bigg\{\frac{32 }{2835 \left(t^2-3\right)^6 \left(t^2-1\right)} \bigg(76265 t^{12}+1803550 t^{10}-11498697 t^8+23078148 t^6\notag\\
&-22718457 t^4-14652738 t^2+23367609\bigg) -\frac{32 }{8505 t \left(t^2-3\right)^7 \left(t^2-1\right)^2}\bigg(433195 t^{18}\notag\\
&+10991120 t^{16}-116769772 t^{14}+420616912 t^{12}-881234862 t^{10}+1592484336 t^8\notag\\
&-2208930444 t^6+1630223280 t^4-375963525 t^2-78382080\bigg) \bigg(\pi -3 \tan ^{-1}(t)\bigg)\notag\\
&-\frac{16 }{35 \left(t^2-3\right)^7 \left(t^2-1\right)^2}\bigg(3465 t^{16}+100380 t^{14}-927360 t^{12}+3217900 t^{10}-7212422 t^8\notag\\
&+14195412 t^6-23299672 t^4+21526436 t^2-7577259\bigg) \log \bigg(\frac{t^2+1}{4} \bigg)+\mathcal{T}^{(C_F)}_2(t)
\bigg\}\notag\\
&+C_FC_A\bigg\{ \frac{8 }{2835 \left(t^2-3\right)^6 \left(t^2-1\right)^3} \bigg(872935 t^{16}+7567140 t^{14}-75054672 t^{12}\notag\\
&+137782532 t^{10}+150426246 t^8-446621748 t^6+8482248 t^4+415400076 t^2\notag\\
&-196677477\bigg) -\frac{8 }{8505 t \left(t^2-3\right)^7 \left(t^2-1\right)^4}\bigg(5275445 t^{22}+30784950 t^{20}-509147387 t^{18}\notag\\
&+1420469896 t^{16}+951601578 t^{14}-10595367292 t^{12}+19702929138 t^{10}-16474375800 t^8\notag\\
&+6298931457 t^6-1215264330 t^4+601890345 t^2-191600640\bigg) \bigg(\pi -3 \tan ^{-1}(t)\bigg)\notag\\
&-\frac{4 }{35 \left(t^2-3\right)^7 \left(t^2-1\right)^4}\bigg(43575 t^{20}+301350 t^{18}-3542245 t^{16}+4989320 t^{14}\notag\\
&+26927838 t^{12}-101446044 t^{10}+157979838 t^8-163463672 t^6+134482283 t^4\notag\\
&-74242330 t^2+17862567\bigg) \log \bigg(\frac{t^2+1}{4} \bigg)+ \mathcal{T}^{(C_A)}_2(t)\bigg\}\,,
\end{align}
\begin{align}
\mathcal{T}^{(C_F)}_2(t)&=\frac{6144 }{t \left(t^2-3\right)^7 \left(t^2+1\right)^2}\bigg(3 t^{14}-42 t^{12}+625 t^{10}-2780 t^8+6613 t^6-5562 t^4\notag\\
&+2647 t^2-160\bigg)g_7^{(2)}+\frac{6144   }{t \left(t^2-3\right)^7 \left(t^2+1\right)^2}\bigg(t^{14}-20 t^{12}+237 t^{10}\notag\\
&-976 t^8+2187 t^6-1916 t^4+775 t^2-32\bigg)g_8^{(2)}+\frac{12 }{\left(t^2-3\right)^7 \left(t^2-1\right)^3 \left(t^2+1\right)^2}\notag\\
&\bigg(135 t^{24}+3668 t^{22}-32786 t^{20}+87876 t^{18}-70871 t^{16}-314200 t^{14}+2461060 t^{12}\notag\\
&-7605976 t^{10}+12371529 t^8-11499772 t^6+6054766 t^4-1628780 t^2+202023\bigg) g_9^{(2)}\notag\\
&-\frac{96 }{\left(t^2-3\right)^7 \left(t^2-1\right)^3 \left(t^2+1\right)^2} \bigg(17 t^{24}+452 t^{22}-3974 t^{20}+9700 t^{18}-569 t^{16}-74936 t^{14}\notag\\
&+412748 t^{12}-1163480 t^{10}+1835439 t^8-1686668 t^6+878042 t^4-228748 t^2+26073\bigg) g_4^{(2)}\notag\\
&+\frac{24 }{\left(t^2-3\right)^7 \left(t^2-1\right)^3 \left(t^2+1\right)^2}\bigg(135 t^{24}+3668 t^{22}-32786 t^{20}+87876 t^{18}-70871 t^{16}\notag\\
&-301912 t^{14}+2268548 t^{12}-6889176 t^{10}+11146825 t^8-10348796 t^6+5411694 t^4\notag\\
&-1411692 t^2+165159\bigg) \tan ^{-1}(t)^2-\frac{128}{81 \sqrt{3} \left(t^2-3\right)^7}\bigg(2 t^{14}-129 t^{12}+3780 t^{10}-8559 t^8\notag\\
&+127170 t^6+838593 t^4-682344 t^2-111537\bigg)g_1^{(2)}-\frac{12}{\left(t^2-3\right)^5 \left(t^2-1\right)^5 \left(t^2+1\right)}\notag\\
&\bigg(33 t^{18}+1121 t^{16}-3436 t^{14}+8836 t^{12}-23058 t^{10}+36830 t^8-34140 t^6+18036 t^4\notag\\
&-4679 t^2+969\bigg)g_2^{(2)}+\frac{1179648 \left(t^{10}-10 t^8+23 t^6-17 t^4+12 t^2-1\right)}{t \left(t^2-3\right)^7 \left(t^2+1\right)^2}g_6^{(2)} \notag\\
&-\frac{18432 (t-1) t (t+1) \left(t^{10}-19 t^8+186 t^6-470 t^4+981 t^2-391\right) }{\left(t^2-3\right)^7 \left(t^2+1\right)^2}g_5^{(2)} \,,
\end{align}
\begin{align}
\mathcal{T}^{(C_A)}_2(t)&=\frac{24576\left(t^{12}-23 t^{10}+266 t^8-930 t^6+681 t^4-423 t^2-20\right)}{t \left(t^2-3\right)^7 \left(t^2+1\right)^2}g_7^{(2)}\notag\\
&+\frac{12288  \left(t^{12}-21 t^{10}+178 t^8-586 t^6+501 t^4-257 t^2-8\right) }{t \left(t^2-3\right)^7 \left(t^2+1\right)^2}g_8^{(2)}\notag\\
&+\frac{3}{\left(t^2-3\right)^7 \left(t^2-1\right)^5 \left(t^2+1\right)^2}\bigg(1657 t^{28}+11642 t^{26}-143785 t^{24}+180852 t^{22}\notag\\
&+1216825 t^{20}-3297130 t^{18}-3018817 t^{16}+27516824 t^{14}-61945741 t^{12}\notag\\
&+84100358 t^{10}-80819475 t^8+55150516 t^6-23111589 t^4+4051754 t^2-8579\bigg) g_9^{(2)}\notag\\
&-\frac{24 }{\left(t^2-3\right)^7 \left(t^2-1\right)^5 \left(t^2+1\right)^2} \bigg(207 t^{28}+1462 t^{26}-18111 t^{24}+24172 t^{22}\notag\\
&+140623 t^{20}-353478 t^{18}-595239 t^{16}+4040872 t^{14}-8983259 t^{12}+12403402 t^{10}\notag\\
&-12167013 t^8+8407340 t^6-3546051 t^4+633862 t^2-5173\bigg) g_4^{(2)}\notag\\
&+\frac{6}{\left(t^2-3\right)^7 \left(t^2-1\right)^5 \left(t^2+1\right)^2}\bigg(1657 t^{28}+11642 t^{26}-143785 t^{24}+180852 t^{22}\notag\\
&+1216825 t^{20}-3305322 t^{18}-2666561 t^{16}+25419672 t^{14}-56244109 t^{12}+75236614 t^{10}\notag\\
&-72250643 t^8+49842100 t^6-21014437 t^4+3552042 t^2+48765\bigg)( \tan ^{-1}t)^2\notag\\
&-\frac{64 }{81 \sqrt{3} \left(t^2-3\right)^7}\bigg(11 t^{14}-903 t^{12}+16335 t^{10}-142155 t^8+795825 t^6-2544453 t^4\notag\\
&+44469 t^2+500823\bigg)g_1^{(2)}-\frac{3}{\left(t^2+1\right) \left(t^4-4 t^2+3\right)^5}\bigg(415 t^{22}+4945 t^{20}-11639 t^{18}\notag\\
&-40945 t^{16}+158774 t^{14}-162998 t^{12}-17934 t^{10}+144798 t^8-99781 t^6+44309 t^4\notag\\
&-30859 t^2+8867\bigg)g_2^{(2)}-\frac{36864 t \left(t^{10}-17 t^8+98 t^6-394 t^4+381 t^2-133\right) }{\left(t^2-3\right)^7 \left(t^2+1\right)^2}g_5^{(2)}\notag\\
&-\frac{294912 \left(t^{10}-20 t^8+48 t^6-30 t^4+31 t^2+2\right)}{t \left(t^2-3\right)^7 \left(t^2+1\right)^2}g_6^{(2)}\,,
\end{align}
where we need five more function bases
\begin{align}
g_5^{(2)}&=\frac{1}{3} \log \left(3-t^2\right) \bigg( \pi -3 \tan ^{-1}(t)\bigg)+\log \left(t^2+1\right) \tan ^{-1}(t)-D_2^{-}\bigg(\frac{t-\sqrt{3}}{t-i}\bigg)-D_2^{-}\bigg(\frac{t+\sqrt{3}}{t-i}\bigg)\,,\notag\\
g_6^{(2)}&=D_2^{-}\left(i t\right)-\frac{1}{2} \log \left(t^2+1\right) \tan ^{-1}(t)-\frac{1}{3}\log (2 t)\bigg(\pi-3\tan ^{-1}(t)\bigg)\,,\notag\\
g_7^{(2)}&=D_2^{-}\bigg(\frac{t-i}{t+i}\bigg),\quad g_8^{(2)}=\pi \log \left(t^2+1\right),\quad g_9^{(2)}=\zeta_2\,.
\end{align}

\begin{figure}[htb]
    \centering
    \includegraphics[scale=1.0]{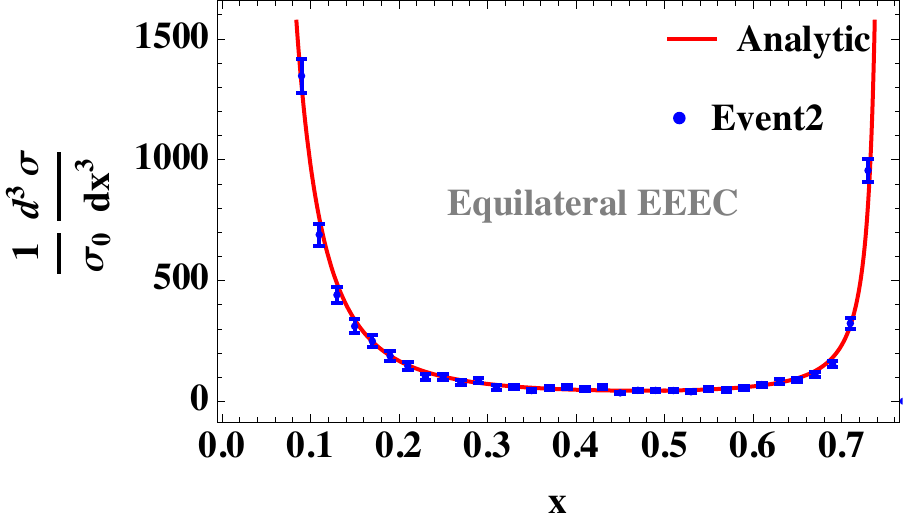}
    \caption{The comparison between the analytic expression and \texttt{Event2} for equilateral EEEC. We compute 2.5 billion events and set the internal cutoff to $10^{-14}$.}
    \label{fig:event2_equ_eeec}
\end{figure}

In Fig.~\ref{fig:event2_equ_eeec}, we also show the equilateral EEEC result from \texttt{Event2}, which has good agreement with our analytic formula.

Even slicing the kinematic space with equilateral constraint gives us an interesting result. There are two singular limits: the collinear limit $x\to 0$ and coplanar limit $x\to\frac{3}{4}$. The collinear expansion is given by
\beq
\frac{1}{\sigma_{\text{tot}}}\frac{d^3\sigma}{dx^3}\overset{x\to0}{\approx}\bigg(\frac{\alpha_s}{4\pi}\bigg)^2\frac{1}{4\pi x \sqrt{3-4x}}\times\bigg(\frac{1}{x^2}\mathcal{F}_{1}+\frac{1}{x^1}\mathcal{F}_{2}+\mathcal{F}_{3}+x \mathcal{F}_{4}+\mathcal{O}(x^2)\bigg)\,,
\label{eq:smallx}
\eeq
where the coefficients are linear combinations of Clausen functions and Riemann zeta functions:
\begin{align}
\mathcal{F}_{1}&=C_F n_f T_F\left(\frac{1856 \kappa }{27
   \sqrt{3}}-\frac{5354}{135}\right) +C_F^2 \left(24
   \zeta _2-\frac{32 \kappa }{27 \sqrt{3}}-\frac{4543}{135}\right)\notag\\
   &\hspace{4.9cm}+C_A C_F \left(-12 \zeta _2+\frac{304 \kappa }{27
   \sqrt{3}}+\frac{779}{54}\right) ,\notag\\
   \mathcal{F}_2&=C_F n_f T_F\left(\frac{4511}{81}-\frac{7
   552 \kappa }{81 \sqrt{3}}\right) +C_F^2 \left(-24
   \zeta _2+\frac{32 \kappa }{81 \sqrt{3}}+\frac{20506}{405}\right)\notag\\
   &\hspace{4.9cm}+C_A C_F \left(12 \zeta _2+\frac{704 \kappa }{27
   \sqrt{3}}-\frac{18047}{540}\right) ,\notag\\
   \mathcal{F}_3&=C_F n_f T_F\left(\frac{206168}{42525}
   -\frac{1408 \kappa }{243 \sqrt{3}}\right) +C_F^2
   \left(48 \zeta _2-\frac{1312 \kappa }{243
   \sqrt{3}}-\frac{2390434}{42525}\right)\notag\\
   &\hspace{4.9cm}+C_A C_F \left(-24 \zeta _2-\frac{944 \kappa }{81
   \sqrt{3}}+\frac{678556}{14175}\right) ,\notag\\
   \mathcal{F}_4&=C_F n_f
   T_F\frac{79}{42} +C_F^2 \left(48 \zeta _2-\frac{320 \kappa }{243
   \sqrt{3}}-\frac{4063357}{85050}\right)\notag\\
   &\hspace{4.9cm}+C_A C_F \left(-24 \zeta _2-\frac{3632 \kappa }{243
   \sqrt{3}}+\frac{16887929}{340200}\right) .
\label{eq:smalleeec}
\end{align}
Here  $\kappa\equiv \text{Cl}_2\left(\frac{\pi}{3}\right)=\text{Im Li}_2 e^{\frac{i\pi}{3}}$ is Gieseking’s constant, with $\text{Cl}_2(\phi)=-\int_{0}^{\phi} \log |2 \sin\frac{x}{2} | dx$ being the Clausen function. This is a transcendentality-two number that is typical in the trijet computation (e.g., the one-loop trijet soft function \cite{Bhattacharya:2022dtm}). Another interesting feature is from the $n_f$ color factor, i.e., the coefficient of $n_f$ for $\mathcal{F}_4$ in Eq.~\eqref{eq:smalleeec} doesn't involve $\kappa$ while $\kappa$ still shows up for $\mathcal{F}_1, \mathcal{F}_2, \mathcal{F}_3$. We will find a similar feature in the triple collinear limit in the next subsection.

Due to the kinematic cut $\Theta\left(\frac{3}{4}-x\right)$, there is no back-to-back limit in the equilateral configuration. Instead, the three detectors are separated by an angle $\frac{2\pi}{3}$ on the same plane, which refers to the \textit{coplanar} limit. The coplanar expansion includes both fractional power divergence and logarithmic divergence: 
\begin{multline}
\frac{1}{\sigma_0}\frac{d^3\sigma}{dx^3}\overset{x\to\frac{3}{4}}{\approx}\bigg(\frac{\alpha_s}{4\pi}\bigg)^2\frac{1}{4\pi}\times\bigg(\frac{\ln(\frac{3}{4}- x)}{\frac{3}{4}-x}\mathcal{R}_1+\frac{1}{\frac{3}{4}-x}\mathcal{R}_2 + \frac{1}{\sqrt{\frac{3}{4}-x}}\mathcal{R}_3\\
+\ln(\frac{3}{4}- x)\mathcal{R}_4+\mathcal{R}_5\bigg)+\mathcal{O}\left(\sqrt{\frac{3}{4}-x} \right)\,,
\end{multline}
and the corresponding coefficients are
\begin{align}
\mathcal{R}_1 &= -\pi \frac{16384}{2187}C_F\left(C_A+2C_F\right) \,, \notag\\
\mathcal{R}_2 & =C_F T_f n_f  \frac{32768}{6561}\pi- C_F^2\bigg(\frac{16384}{729}\pi +\frac{131072}{2187}\pi \ln 2 \bigg) -C_F C_A\bigg( \frac{90112}{6561}\pi +\frac{65536}{2187}\pi \ln 2 \bigg)\,,\notag\\
\mathcal{R}_3 &= C_F T_f n_f \bigg( -\frac{3200}{243 \sqrt{3}}\kappa-\frac{2062 }{81}\text{Li}_2(-3) +\frac{3874}{81}\zeta_2-\frac{339824}{2835}-\frac{5584528 }{25515}\ln 2\bigg)\notag\\
&+C_F^2\bigg( \frac{2560}{729 \sqrt{3}}\kappa -\frac{4976 }{243}\text{Li}_2(-3)+\frac{892816}{2187}\zeta_2+\frac{611200}{5103}+\frac{5170304}{5103}\ln 2 \bigg)\notag\\
&+C_F C_A\bigg( -\frac{58240}{729 \sqrt{3}}\kappa -\frac{8317 }{243}\text{Li}_2(-3) -\frac{50941}{2187}\zeta_2+\frac{3939608}{8505}-\frac{55173784}{76545}\ln 2 \bigg)\,,\notag\\
\mathcal{R}_4&=\pi \bigg( C_F^2 \frac{4653056}{6561}-C_F C_A \frac{3276800}{6561} \bigg)\,,\notag\\
\mathcal{R}_5&=C_F T_f n_f \frac{5079040}{19683}\pi+C_F^2\bigg( -\frac{376832}{243}\pi+\frac{18612224}{6561}\pi  \ln 2+\frac{1576960}{2187}\pi \ln 3 \bigg)\notag\\
&+C_F C_A\bigg( -\frac{2097152}{19683}\pi-\frac{13107200}{6561}\pi \ln 2-\frac{1077248 }{2187}\pi \ln 3 \bigg)\,, 
\end{align}
where we need one more transcendentality-two number $\text{Li}_2(-3)$. The $C_A+2C_F$ structure in $\mathcal{R}_1$ implies that the leading logarithm in the cumulant can possibly be predicted by a Sudakov form factor \cite{Sudakov:1954sw}.

It would be interesting to study the singularity structure in both limits and resum the large logarithms in the future. In the collinear limit, our result provides the regular terms that complete the two-loop equilateral EEEC jet function. 
To recover its close form in $\epsilon$, one might need to compute equilateral EEEC to higher orders in $\epsilon$ expansion. The soft gluon enhancement appears in the coplanar limit, and similarly, our fixed-order calculation provides the needed ingredients for its resummation. Some of the similar analysis for another trijet event shape observable $D$-parameter can be found in Refs.~\cite{Banfi:2000ut,Banfi:2001pb,Larkoski:2018cke}.
Either way, equilateral EEEC contains valuable information on understanding the symmetric trijet events in electron-positron collisions.

Like two-point energy correlator (EEC), equilateral EEEC has nice analytic properties and is free of \textit{Sudakov shoulders} \cite{Catani:1997xc}. For event shape observables like thrust, $C$ parameter and heavy jet mass, the range of the parameter grows order by order in perturbation theory, and the incomplete cancellation between real emissions and virtual corrections leads to divergences or kinks at fixed orders. To obtain a precise measurement of the strong running coupling $\alpha_s$, one will have to resum the Sudakov shoulder logarithms that fall into the relevant regions \cite{Bhattacharya:2022dtm}. However, in equilateral EEEC, since the three particles are separated by the same angle, the maximum angle is $\frac{2\pi}{3}$ when all three particles fall into the same plane. This geometry constraint remains the same in higher-order perturbation theory so that the IR cancellation is guaranteed by Kinoshita-Lee-Nauenberg (KLN) theorem \cite{Kinoshita:1962ur,PhysRev.133.B1549}.

\subsection{Triple collinear limit at next-to-leading power and beyond}
\label{subsec:TripleCollinear}
The factorization theoroem at leading power (LP) has been well understood for different observables and different physical processes. It allows for the resummation of large logarithms to very high accuracy. Oppositely, much less is known for the factorization theorem and its violation at next-to-leading power (NLP). The complete factorization framework is still not established for NLP observables. In this subsection, we focus on the NLP contribution from the direct calculation point of view, where only a few cases have been carried out~\cite{Balitsky:2017gis,Ebert:2018gsn,Moult:2019vou,Cieri:2019tfv,Ebert:2020dfc,Moos:2020wvd,Oleari:2020wvt}. More discussions can be found in Ref.~\cite{Ebert:2021jhy}.

At LP, the triple collinear EEEC is factorized as the hard function $\vec{H}=\{H_q,H_g\}$ and the jet function $\vec{J}=\{J_q,J_g\}$, which both live in the flavor space. In momentum space, the EEEC jet function also decouples into the triple collinear phase space~\cite{Gehrmann-DeRidder:1997fom,Ritzmann:2014mka} and the $1\to 3$ splitting functions~\cite{Campbell:1997hg,Catani:1998nv,Ritzmann:2014mka}. Benefiting from the decoupling, the calculation was performed in Ref.~\cite{Chen:2019bpb}, which becomes the first analytic calculation of a three-parameter jet substructure observable. From our EEEC result with full angle dependence in Section~\ref{sec:results}, it is straightforward to extract NLP contribution in the triple collinear limit. While the NLP contribution itself can provide comparison data for the study of NLP factorization, it can not provide hints toward NLP factorization. It is more interesting to explore a similar decoupling of the phase space and the integrand to NLP, and extract the NLP corrections from a direct computation. 

For the purpose of extracting the triple collinear limit, we perform the following rescaling
\begin{equation}
x_1 \to \lambda \, x_1 , \quad x_2 \to \lambda \, x_2 , \quad x_3 \to \lambda \, x_3 
\end{equation} 
and expand the corresponding formula in $\lambda$ order by order. To decouple the phase space measure and the integrand to NLP, we start from Eq.~\eqref{eq:boundz1z2} and reformulate it as in the following, 
\begin{equation}
\label{eq:z1z2BoundG}
\frac{1}{\sigma_{\text{tot}}}\frac{d^3\sigma}{dx_1dx_2dx_3} = \int_0^1 dz_1 \int_0^{\frac{1-z_1}{1- x_3 z_1}} dz_2 \, g\left(x_1,x_2,x_3,z_1,z_2 \right)\,,
\end{equation}
where $g\left(x_1,x_2,x_3,z_1,z_2 \right)$ is used to represent the EEEC integrand. Notice that the upper bound of $z_2$ depends on $x_3$, this makes the decoupling non-trivial at NLP. At LP, it is safe to expand the upper bound and the integrand separately, where the leading terms in $\lambda$  directly gives us the decoupling at LP, as computed in Ref.~\cite{Chen:2019bpb}. At NLP, one can not just expand the integrand $g\left(x_1,x_2,x_3,z_1,z_2 \right)$ to the next-to-leading term while only keeping the leading term in the upper bound. One may expand both the upper bound and the integrand to next-to-leading terms, however, it doesn't make the computation simpler and also mixes a part of NNLP and beyond into the NLP contribution. 

To extract the exact NLP contribution and make the decoupling explicit, we separate the interval of $z_2$ integration into the following three intervals, 
\begin{equation}
\label{eq:decomposition}
\int_0^{\frac{1-z_1}{1- x_3 z_1}} dz_2 = \int_0^{1-z_1} dz_2  +  \int_{1-z_1}^{(1-z_1)(1+x_3 z_1)} dz_2  + \int^{{\frac{1-z_1}{1- x_3 z_1}} }_{(1-z_1)(1+x_3 z_1)} dz_2  
\end{equation} 
On the right-hand side of Eq.~\eqref{eq:decomposition}, the first term corresponds to the triple collinear phase space measure and contributes to LP and beyond, the second term starts to contribute at NLP, and the third term only contributes to NNLP and beyond. The right-hand side of Eq.~\eqref{eq:decomposition} is formulated in a way that the $i$-th term contributes only at N$^{i-1}$LP and beyond. It is similar to Ref.~\cite{Gehrmann:2022euk} where the operators are organized in a way such that the $i$-th type operators contribute only at $\alpha_s^i$ and beyond. 
Let us focus on the second term, together with the integrand, we have 
\begin{align}
 &\hspace{0.5cm}\int_{1-z_1}^{(1-z_1)(1+x_3 z_1)}  dz_2 \, g\left(x_1,x_2,x_3,z_1,z_2 \right) \notag  \\ 
  &=  (1-z_1) \int_0^{x_3 z_1} d t \, g\big( x_1,x_2,x_3,z_1,z_2=(1-z_1)(1+t) \big) \notag   \\
   &=  x_3 z_1 (1-z_1) \, g(x_1,x_2,x_3,z_1,z_2=1-z_1) +\cdots \notag   \\
   &=  x_3 z_1 (1-z_1) \int_0^{1-z_1} dz_2  \delta\left(z_2- (1-z_1)\right)\, g(x_1,x_2,x_3,z_1,z_2) +\cdots \,,
\end{align} 
%|_{\text{leading term in }\lambda} + \cdots
where $\cdots$ only contributes to NNLP and beyond. 
In summary, the contribution up to NLP can be written as
\begin{multline}
\label{eq:toNLPformula}
\frac{1}{\sigma_\text{tot}}\frac{d^3\sigma}{dx_1 dx_2 dx_3}\overset{\text{triple coll}}{\approx} \int_0^1 dz_1 \int_0^{1-z_1} dz_2 \, \bigg[  g(x_1,x_2,x_3,z_1,z_2)  \\
\times \bigg( 1+ \red{x_3 z_1 (1-z_1) \delta\big( z_2- (1-z_1)\big)}  \bigg) \bigg] + \mathcal{O}(\text{NNLP}) \,. 
\end{multline}
%\begin{align}
%\label{eq:toNLPformula}
%\frac{1}{\sigma_\text{tot}}&\frac{d^3\sigma}{dx_1 dx_2 dx_3}\overset{\text{triple coll}}{\approx} \int_0^1 dz_1 \int_0^{1-z_1} dz_2  \left[ g(x_1,x_2,x_3,z_1,z_2)\right]|_{\text{leading term in }\lambda} \nonumber \\
%&+ \bigg( \int_0^1 dz_1 \int_0^{1-z_1} dz_2 \left[ g(x_1,x_2,x_3,z_1,z_2)\right]|_{\text{next-to-leading term in }\lambda} \nonumber \\ 
%&+  \int^1_0 dz_1 x_3 z_1 (1-z_1) \left[ g(x_1,x_2,x_3,z_1,z_2=1-z_1)\right]|_{\text{leading term in }\lambda} \bigg)+ \mathcal{O}(\text{NNLP}) \,. 
%\end{align} 
The intervals of both integration variables $z_1\,, z_2$ in the above equation don't involve any kinematic variables $x_{1,2,3}$. Therefore, we can safely expand the integrand in the triple collinear limit, which makes the computation of the NLP contribution as easy as LP. We emphasize that the contact term that is proportional to $\red{\delta\big( z_2 - (1-z_1) \big)}$ is crucial to get correct result at NLP. The above method to extract NLP for triple collinear EEEC may also be useful to compute the NLP contributions for other observables. It is also straightforward to generalize the above method to NNLP and beyond.  

An alternative method to extract the NLP contribution is to first integrate over $z_2$ in Eq.~\eqref{eq:z1z2BoundG} without performing any expansion. By simplifying the resulting integrand and performing the expansion, one can integrate over $z_1$ and obtain the NLP contribution. The simplified integrand with $z_1$ dependence is also useful as a cross-check of our final analytic formula, such that we also provide it in the ancillary file.       

We use both methods to extract the triple collinear limit to NLP and find the same result. The validity of both methods is verified by the fact that no poles in the integration variables are generated when performing the expansion. We also compare the result with the final full analytic formula by setting $x_{1,2,3}$ to very small numbers, and we find the difference is indeed an NNLP contribution.    

The result up to NLP in the triple collinear limit can be written as 
\begin{equation}
\label{eq:tripleCollinearF}
\frac{1}{\sigma_\text{tot}}\frac{d^3\sigma}{dx_1 dx_2 dx_3}\overset{\text{triple coll}}{\approx} \left(\frac{\alpha_s}{4\pi}\right)^2\frac{1}{4\pi\sqrt{-s_2^2}}\left(\frac{C^{\text{LP}}(x_i)}{\lambda^2}+\frac{C^{\text{NLP}}(x_i)}{\lambda^1}+\mathcal{O}(\lambda^0)\right)\,, 
\end{equation}
where $\lambda$ is just used to track the expansion order and should be set to 1 at the end. The leading contribution $C^{\text{LP}}$ agrees with the result in Ref.~\cite{Chen:2019bpb}. The subleading term $C^{\text{NLP}}$ is new and also contains three color channels:
\begin{equation}
    C^{\text{NLP}}(x_i)=C_F T_F n_f A_{n_f}(x_i)+C_F^2 A_{C_F}(x_i)+C_F C_A A_{C_A}(x_i)
\end{equation}
The $n_f$ contribution is given below:
\begin{align}
    A_{n_f}&=\frac{1}{s_1^{10}}\bigg[ -\frac{16 x_1^{13}}{x_3^4}+\left(\frac{176
   x_2}{x_3^4}+\frac{140}{x_3^3}\right)
   x_1^{12}+\left(-\frac{864 x_2^2}{x_3^4}-\frac{1024
   x_2}{x_3^3}-\frac{23672}{45 x_3^2}\right)
   x_1^{11}\notag\\
   &+\left(\frac{2464 x_2^3}{x_3^4}+\frac{2680
   x_2^2}{x_3^3}+\frac{6040 x_2}{3 x_3^2}+\frac{9703}{9
   x_3}\right) x_1^{10}+\left(-\frac{4400
   x_2^4}{x_3^4}-\frac{1280 x_2^3}{x_3^3}+\frac{28616
   x_2^2}{45 x_3^2}\right.\notag\\
   &\left.-\frac{10382 x_2}{45
   x_3}-\frac{108919}{90}\right) x_1^9+\left(\frac{4752
   x_2^5}{x_3^4}-\frac{9420 x_2^4}{x_3^3}-\frac{14968
   x_2^3}{x_3^2}-\frac{64177 x_2^2}{5 x_3}-\frac{47843
   x_2}{9}\right) x_1^8\notag\\
   &+\left(-\frac{2112
   x_2^6}{x_3^4}+\frac{26880 x_2^5}{x_3^3}+\frac{432688
   x_2^4}{15 x_3^2}+\frac{415816 x_2^3}{15 x_3}+\frac{111187
   x_2^2}{5}+\frac{172757 x_3 x_2}{15}\right)
   x_1^7\notag\\
   &+\left(-\frac{17976 x_2^6}{x_3^3}-\frac{240016
   x_2^5}{15 x_3^2}-\frac{279406 x_2^4}{15 x_3}-\frac{858662
   x_2^3}{45}-\frac{258220}{9} x_3 x_2^2\right)
   x_1^6\notag\\
   &+\left(\frac{2894 x_2^5}{x_3}+\frac{35138
   x_2^4}{9}+\frac{559376}{45} x_3 x_2^3+\frac{27734}{5} x_3^2
   x_2^2\right) x_1^5+\left(-\frac{35887}{15} x_3
   x_2^4-\frac{18322}{5} x_3^2 x_2^3\right)x_1^4\notag\\
   &+\frac{110872}{45} x_2^3 x_3^3 x_1^3 \bigg]+\frac{1}{s_1^{12}}\red{\log(x_1)}\bigg[-\frac{16 x_1^{15}}{x_3^4}+\left(\frac{192
   x_2}{x_3^4}+\frac{168}{x_3^3}\right)
   x_1^{14}+\left(-\frac{1040 x_2^2}{x_3^4}\right.\notag\\
   &\left.-\frac{1424
   x_2}{x_3^3}-\frac{2344}{3 x_3^2}\right)
   x_1^{13}+\left(\frac{3328 x_2^3}{x_3^4}+\frac{4736
   x_2^2}{x_3^3}+\frac{12104 x_2}{3 x_3^2}+\frac{6260}{3
   x_3}\right) x_1^{12}\notag\\
   &+\left(-\frac{6864
   x_2^4}{x_3^4}-\frac{6160 x_2^3}{x_3^3}-\frac{20136 x_2^2}{5
   x_3^2}-\frac{11896 x_2}{3 x_3}-\frac{17076}{5}\right)
   x_1^{11}+\left(\frac{9152 x_2^5}{x_3^4}-\frac{5720
   x_2^4}{x_3^3}\right.\notag\\
   &\left.-\frac{85496 x_2^3}{5 x_3^2}-\frac{49244
   x_2^2}{3 x_3}-\frac{8572 x_2}{3}\right)
   x_1^{10}+\left(-\frac{6864 x_2^6}{x_3^4}+\frac{35904
   x_2^5}{x_3^3}+\frac{280784 x_2^4}{5 x_3^2}\right.\notag\\
   &\left.+\frac{60352
   x_2^3}{x_3}-\frac{16\left(-44693 x_2^2-23320 x_3
   x_2\right)}{15} \right) x_1^9+\left(-\frac{65472
   x_2^6}{x_3^3}-\frac{315216 x_2^5}{5 x_3^2}\right.\notag\\
   &\left.-\frac{198616
   x_2^4}{3 x_3}-\frac{4\left(246961 x_2^3+345869 x_3
   x_2^2\right)}{15} \right) x_1^8+\left(\frac{6864
   x_2^8}{x_3^4}+\frac{66528 x_2^7}{x_3^3}+\frac{44464
   x_2^6}{5 x_3^2}\right.\notag\\
   &\left.+\frac{63280 x_2^5}{3 x_3}+\frac{4\left(97897 x_2^4+216666 x_3 x_2^3+92975 x_3^2
   x_2^2\right)}{15}
   \right) x_1^7+\left(-\frac{9152
   x_2^9}{x_3^4}-\frac{38280 x_2^8}{x_3^3}\right.\notag\\
   &\left.+\frac{249616
   x_2^7}{5 x_3^2}-\frac{12488 x_2^6}{x_3}-\frac{8\left(10292 x_2^5+40165 x_3 x_2^4-35663 x_3^2
   x_2^3\right)}{15}
   \right) x_1^6\notag\\
   &+\bigg(\frac{6864
   x_2^{10}}{x_3^4}+\frac{7920 x_2^9}{x_3^3}-\frac{259704
   x_2^8}{5 x_3^2}+\frac{152864 x_2^7}{3 x_3}+\frac{8}{15}
   \left(-14491 x_2^6-4296 x_3 x_2^5\right.\notag\\
   &\left.-21392 x_3^2 x_2^4-24618
   x_3^3 x_2^3\right)\bigg) x_1^5+\left(-\frac{3328
   x_2^{11}}{x_3^4}+\frac{4928 x_2^{10}}{x_3^3}+\frac{272648
   x_2^9}{15 x_3^2}-\frac{158876 x_2^8}{3 x_3}\right.\notag\\
   &\left.-\frac{4\left(-165413 x_2^7-43150 x_3 x_2^6+27668 x_3^2 x_2^5-41818
   x_3^3 x_2^4\right)}{15}
   \right) x_1^4+\bigg(\frac{1040
   x_2^{12}}{x_3^4}\notag\\
   &-\frac{4304 x_2^{11}}{x_3^3}+\frac{36968
   x_2^{10}}{15 x_3^2}+\frac{50056 x_2^9}{3 x_3}+\frac{8}{15}
   \left(-74107 x_2^8+7867 x_3 x_2^7-2535 x_3^2 x_2^6\right.\notag\\
   &\left.+24618
   x_3^3 x_2^5-20909 x_3^4 x_2^4\right)\bigg)
   x_1^3+\left(-\frac{192 x_2^{13}}{x_3^4}+\frac{1336
   x_2^{12}}{x_3^3}-\frac{17304 x_2^{11}}{5 x_3^2}+\frac{2620
   x_2^{10}}{x_3}\right.\notag\\
   &\left.-\frac{4\left(-28243 x_2^9-59399 x_3
   x_2^8+92975 x_3^2 x_2^7+66256 x_3^3 x_2^6-70452 x_3^4
   x_2^5\right)}{15} \right) x_1^2\notag\\
   &+\bigg(\frac{16
   x_2^{14}}{x_3^4}-\frac{160 x_2^{13}}{x_3^3}+\frac{704
   x_2^{12}}{x_3^2}-\frac{1760 x_2^{11}}{x_3}+\frac{8}{15}
   \left(-1031 x_2^{10}-46640 x_3 x_2^9+143235 x_3^2
   x_2^8\right.\notag\\
   &\left.-116200 x_3^3 x_2^7+18590 x_3^4 x_2^6+4296 x_3^5
   x_2^5\right)\bigg) x_1+\frac{17076
   x_2^{11}}{5}+\frac{66088}{5} x_2^6 x_3^5-70216 x_2^7
   x_3^4\notag\\
   &+105380 x_2^8 x_3^3-55204 x_2^9 x_3^2+\frac{17036}{5}
   x_2^{10} x_3+\frac{1}{x_1}\left(-\frac{6260 x_2^{12}}{3}+\frac{17176}{3}
   x_3 x_2^{11}\right.\notag\\
   &\left.+\frac{41384}{3} x_3^2
   x_2^{10}-\frac{231112}{3} x_3^3 x_2^9+119164 x_3^4
   x_2^8-72048 x_3^5 x_2^7+12488 x_3^6
   x_2^6\right)\notag\\
   &+\frac{1}{x_1^2}\left(\frac{2344 x_2^{13}}{3}-\frac{14216}{3}
   x_3 x_2^{12}+7488 x_3^2 x_2^{11}+\frac{43904}{3} x_3^3
   x_2^{10}-\frac{223000}{3} x_3^4 x_2^9\right.\notag\\
   &\left.+114984 x_3^5
   x_2^8-58816 x_3^6 x_2^7\right)+\frac{1}{x_1^3}\left(-168 x_2^{14}+1584
   x_3 x_2^{13}-6072 x_3^2 x_2^{12}+10464 x_3^3 x_2^{11}\right.\notag\\
   &\left.+792
   x_3^4 x_2^{10}-43824 x_3^5 x_2^9+103752 x_3^6 x_2^8-66528
   x_3^7 x_2^7\right)+\frac{1}{x_1^4}\left(16 x_2^{15}-208 x_3 x_2^{14}\right.\notag\\
   &\left.+1232
   x_3^2 x_2^{13}-4368 x_3^3 x_2^{12}+10192 x_3^4
   x_2^{11}-16016 x_3^5 x_2^{10}+16016 x_3^6 x_2^9-6864 x_3^7
   x_2^8\right)\bigg]\notag\\
   &+\frac{1}{s_{1}^{12}}\red{b_2}\bigg[-\frac{16 x_1^{17}}{x_3^5}+\left(\frac{240
   x_2}{x_3^5}+\frac{192}{x_3^4}\right)
   x_1^{16}+\left(-\frac{1664 x_2^2}{x_3^5}-\frac{2208
   x_2}{x_3^4}-\frac{1040}{x_3^3}\right)
   x_1^{15}\notag\\
   &+\left(\frac{7040 x_2^3}{x_3^5}+\frac{10944
   x_2^2}{x_3^4}+\frac{8528
   x_2}{x_3^3}+\frac{3328}{x_3^2}\right)
   x_1^{14}+\left(-\frac{20160 x_2^4}{x_3^5}-\frac{28896
   x_2^3}{x_3^4}-\frac{25168 x_2^2}{x_3^3}\right.\notag\\
   &\left.-\frac{16640
   x_2}{x_3^2}-\frac{6864}{x_3}\right)
   x_1^{13}+\left(\frac{40768 x_2^5}{x_3^5}+\frac{34944
   x_2^4}{x_3^4}+\frac{14352 x_2^3}{x_3^3}+\frac{7488
   x_2^2}{x_3^2}+\frac{11440 x_2}{x_3}\right.\notag\\
   &\left.+9152\right)
   x_1^{12}+\left(-\frac{58240 x_2^6}{x_3^5}+\frac{26208
   x_2^5}{x_3^4}+\frac{106288 x_2^4}{x_3^3}+\frac{124800
   x_2^3}{x_3^2}+\frac{82368 x_2^2}{x_3}\right.\notag\\
   &\left.+16016 x_2\right)
   x_1^{11}+\left(\frac{54912 x_2^7}{x_3^5}-\frac{192192
   x_2^6}{x_3^4}-\frac{331760 x_2^5}{x_3^3}-\frac{375232
   x_2^4}{x_3^2}-\frac{320320 x_2^3}{x_3}\right.\notag\\
   &\left.-201344 x_2^2-75504
   x_3 x_2\right) x_1^{10}+\left(-\frac{22880
   x_2^8}{x_3^5}+\frac{398112 x_2^7}{x_3^4}+\frac{441584
   x_2^6}{x_3^3}+\frac{475904 x_2^5}{x_3^2}\right.\notag\\
   &\left.+\frac{446160
   x_2^4}{x_3}+368384 x_2^3+349984 x_3 x_2^2\right)
   x_1^9+\left(-\frac{247104 x_2^8}{x_3^4}-\frac{212784
   x_2^7}{x_3^3}-\frac{274560 x_2^6}{x_3^2}\right.\notag\\
   &\left.-\frac{267696
   x_2^5}{x_3}-256448 x_2^4-314944 x_3 x_2^3-166784 x_3^2
   x_2^2\right) x_1^8+\left(\frac{54912
   x_2^7}{x_3^2}+\frac{54912 x_2^6}{x_3}\right.\notag\\
   &\left.+62816 x_2^5+119600
   x_3 x_2^4+105024 x_3^2 x_2^3\right) x_1^7+\left(19872 x_3^3
   x_2^3+7168 x_3^2 x_2^4-3328
   x_2^6\right.\notag\\
   &\left.-21008 x_3 x_2^5\right) x_1^6+\left(-1232 x_3^2 x_2^5-32816 x_3^3
   x_2^4\right) x_1^5+9728 x_2^4 x_3^4 x_1^4\bigg]+\bigg[-\frac{16 x_1^4}{x_2^5}\notag\\
   &+\left(\frac{32
   x_3}{x_2^5}-\frac{16}{x_2^4}\right) x_1^3+\left(\frac{48
   x_3}{x_2^4}-\frac{16 x_3^2}{x_2^5}\right) x_1^2-\frac{32
   x_3^2 x_1}{x_2^4}\bigg]\red{b_3(x_3,x_1,x_2)} \nonumber \\
   &+\text{permutations of } x_{1,2,3}\,.
\end{align}
Similarly, for other color factor contributions, we have
\begin{align}
    A_{C_F}&=\frac{1}{s_1^{10}}\bigg[-\frac{4 x_1^{14}}{x_2
   x_3^4}+\left(\frac{52}{x_3^4}+\frac{19}{x_3^3 x_2}\right)
   x_1^{13}+\left(-\frac{308
   x_2}{x_3^4}-\frac{142}{x_3^3}+\frac{365}{9 x_3^2
   x_2}\right) x_1^{12}+\left(\frac{1092
   x_2^2}{x_3^4}\right.\notag\\
   &\left.+\frac{241 x_2}{x_3^3}-\frac{6631}{9
   x_3^2}-\frac{100259}{180 x_3 x_2}\right)
   x_1^{11}+\left(-\frac{2548 x_2^3}{x_3^4}+\frac{1268
   x_2^2}{x_3^3}+\frac{4806 x_2}{x_3^2}+\frac{198539}{30
   x_3}\right) x_1^{10}\notag\\
   &+\left(\frac{4004
   x_2^4}{x_3^4}-\frac{8261 x_2^3}{x_3^3}-\frac{147770
   x_2^2}{9 x_3^2}-\frac{302942 x_2}{15
   x_3}-\frac{210419}{20}\right) x_1^9+\left(-\frac{4004
   x_2^5}{x_3^4}\right.\notag\\
   &\left.+\frac{23342 x_2^4}{x_3^3}+\frac{296575
   x_2^3}{9 x_3^2}+\frac{3200369 x_2^2}{90 x_3}+\frac{346399
   x_2}{10}\right) x_1^8+\left(\frac{1716
   x_2^6}{x_3^4}-\frac{41151 x_2^5}{x_3^3}\right.\notag\\
   &\left.-\frac{38205
   x_2^4}{x_3^2}-\frac{3198389 x_2^3}{90 x_3}-\frac{551761
   x_2^2}{18}-\frac{41662 x_3 x_2}{3}\right)
   x_1^7+\left(\frac{24684 x_2^6}{x_3^3}+\frac{52684 x_2^5}{3
   x_3^2}\right.\notag\\
   &\left.+\frac{255406 x_2^4}{15 x_3}+\frac{64233
   x_2^3}{5}+\frac{105028}{15} x_3 x_2^2\right)
   x_1^6+\left(-\frac{36187 x_2^5}{15 x_3}-\frac{7912
   x_2^4}{5}+\frac{112747}{15} x_3 x_2^3\right.\notag\\
   &\left.+\frac{151714}{45}
   x_3^2 x_2^2\right) x_1^5+\left(-\frac{177793}{30} x_3
   x_2^4-\frac{206644}{45} x_3^2 x_2^3\right)
   x_1^4-\frac{7657}{10} x_2^3 x_3^3 x_1^3\bigg]\notag\\
   &+\frac{1}{s_1^{12}}\red{\log(x_1)}\bigg[-\frac{4 x_1^{16}}{x_2
   x_3^4}+\left(\frac{56}{x_3^4}+\frac{26}{x_3^3 x_2}\right)
   x_1^{15}+\left(-\frac{360
   x_2}{x_3^4}-\frac{240}{x_3^3}+\frac{14}{3 x_3^2 x_2}\right)
   x_1^{14}\notag\\
   &+\left(\frac{1400 x_2^2}{x_3^4}+\frac{778
   x_2}{x_3^3}-\frac{1646}{3 x_3^2}-\frac{1907}{3 x_3
   x_2}\right) x_1^{13}+\left(-\frac{3640
   x_2^3}{x_3^4}-\frac{104 x_2^2}{x_3^3}+\frac{15796 x_2}{3
   x_3^2}\right.\notag\\
   &\left.+\frac{29168}{3 x_3}\right) x_1^{12}+\left(\frac{6552
   x_2^4}{x_3^4}-\frac{7566 x_2^3}{x_3^3}-\frac{69076 x_2^2}{3
   x_3^2}-\frac{187597 x_2}{5 x_3}-21625\right)
   x_1^{11}\notag\\
   &+\left(-\frac{8008 x_2^5}{x_3^4}+\frac{29744
   x_2^4}{x_3^3}+\frac{58014 x_2^3}{x_3^2}+\frac{1300108
   x_2^2}{15 x_3}+\frac{1532353 x_2}{15}\right)
   x_1^{10}+\left(\frac{5720 x_2^6}{x_3^4}\right.\notag\\
   &\left.-\frac{64350
   x_2^5}{x_3^3}-\frac{89166 x_2^4}{x_3^2}-\frac{627674
   x_2^3}{5 x_3}-\frac{2177254 x_2^2}{15}-\frac{1130216 x_3
   x_2}{15}\right) x_1^9+\left(\frac{92664
   x_2^6}{x_3^3}\right.\notag\\
   &\left.+\frac{76824 x_2^5}{x_3^2}+\frac{1706228
   x_2^4}{15 x_3}+\frac{1917257 x_2^3}{15}+\frac{2001659}{15}
   x_3 x_2^2\right) x_1^8+\left(-\frac{5720
   x_2^8}{x_3^4}-\frac{93522 x_2^7}{x_3^3}\right.\notag\\
   &\left.-\frac{12584
   x_2^6}{x_3^2}-\frac{995477 x_2^5}{15 x_3}-\frac{1016974
   x_2^4}{15}-\frac{363547}{5} x_3 x_2^3-37414 x_3^2
   x_2^2\right) x_1^7+\left(\frac{8008
   x_2^9}{x_3^4}\right.\notag\\
   &\left.+\frac{66352 x_2^8}{x_3^3}-\frac{55946
   x_2^7}{x_3^2}+\frac{42512 x_2^6}{x_3}+\frac{238238
   x_2^5}{15}+21916 x_3 x_2^4+\frac{425752}{15} x_3^2
   x_2^3\right) x_1^6\notag\\
   &+\left(-\frac{6552
   x_2^{10}}{x_3^4}-\frac{31746 x_2^9}{x_3^3}+\frac{232606
   x_2^8}{3 x_3^2}-\frac{833968 x_2^7}{15 x_3}+\frac{212812
   x_2^6}{15}-\frac{67118}{15} x_3 x_2^5\right.\notag\\
   &\left.-\frac{254474}{15}
   x_3^2 x_2^4+\frac{23356}{5} x_3^3 x_2^3\right)
   x_1^5+\left(\frac{3640 x_2^{11}}{x_3^4}+\frac{8840
   x_2^{10}}{x_3^3}-\frac{162316 x_2^9}{3 x_3^2}+\frac{915092
   x_2^8}{15 x_3}\right.\notag\\
   &\left.-\frac{437396 x_2^7}{15}+\frac{99122}{15} x_3
   x_2^6+14074 x_3^2 x_2^5-\frac{53716}{5} x_3^3 x_2^4\right)
   x_1^4+\left(-\frac{1400 x_2^{12}}{x_3^4}-\frac{442
   x_2^{11}}{x_3^3}\right.\notag\\
   &\left.+\frac{67724 x_2^{10}}{3
   x_3^2}-\frac{553513 x_2^9}{15 x_3}+\frac{281428
   x_2^8}{15}+\frac{24851}{15} x_3 x_2^7-\frac{51612}{5} x_3^2
   x_2^6-\frac{23356}{5} x_3^3 x_2^5\right.\notag\\
   &\left.+\frac{53716}{5} x_3^4
   x_2^4\right) x_1^3+\left(\frac{360
   x_2^{13}}{x_3^4}-\frac{624 x_2^{12}}{x_3^3}-\frac{16394
   x_2^{11}}{3 x_3^2}+\frac{42284 x_2^{10}}{5
   x_3}+\frac{180259 x_2^9}{15}\right.\notag\\
   &\left.-\frac{554839}{15} x_3
   x_2^8+37414 x_3^2 x_2^7-\frac{270916}{15} x_3^3
   x_2^6+\frac{43364}{15} x_3^4 x_2^5\right)
   x_1^2+\left(-\frac{56 x_2^{14}}{x_3^4}+\frac{214
   x_2^{13}}{x_3^3}\right.\notag\\
   &\left.+\frac{638 x_2^{12}}{x_3^2}+\frac{45986
   x_2^{11}}{15 x_3}-\frac{446368
   x_2^{10}}{15}+\frac{1130216}{15} x_3 x_2^9-\frac{289364}{3}
   x_3^2 x_2^8+\frac{213158}{3} x_3^3 x_2^7\right.\notag\\
   &\left.-\frac{427862}{15}
   x_3^4 x_2^6+\frac{67118}{15} x_3^5 x_2^5\right) x_1+21625
   x_2^{11}-30070 x_2^6 x_3^5+96958 x_2^7 x_3^4-146579 x_2^8
   x_3^3\notag\\
   &+133133 x_2^9 x_3^2-72399 x_2^{10} x_3-\frac{2632
   x_2^{12}}{x_3}-\frac{16 x_2^{13}}{x_3^2}-\frac{24
   x_2^{14}}{x_3^3}+\frac{4 x_2^{15}}{x_3^4}+\frac{1}{x_1}\left(\frac{1907
   x_2^{13}}{3 x_3}\right.\notag\\
   &\left.-\frac{21272 x_2^{12}}{3}+\frac{103361}{3}
   x_3 x_2^{11}-\frac{285392}{3} x_3^2
   x_2^{10}+\frac{487307}{3} x_3^3 x_2^9-\frac{524264}{3}
   x_3^4 x_2^8\right.\notag\\
   &\left.+121963 x_3^5 x_2^7-42512 x_3^6
   x_2^6\right)+\frac{1}{x_1^2}\left(-\frac{14 x_2^{14}}{3 x_3}+\frac{1694
   x_2^{13}}{3}-\frac{17710}{3} x_3 x_2^{12}+28490 x_3^2
   x_2^{11}\right.\notag\\
   &\left.-\frac{241766}{3} x_3^3 x_2^{10}+\frac{429814}{3}
   x_3^4 x_2^9-\frac{463078}{3} x_3^5 x_2^8+68530 x_3^6
   x_2^7\right)+\frac{1}{x_1^3}\bigg(-\frac{26 x_2^{15}}{x_3}\notag\\
   &+264
   x_2^{14}-992 x_3 x_2^{13}+728 x_3^2 x_2^{12}+8008 x_3^3
   x_2^{11}-38584 x_3^4 x_2^{10}+96096 x_3^5 x_2^9\notag\\
   &-159016
   x_3^6 x_2^8+93522 x_3^7 x_2^7\bigg)+\frac{1}{x_1^4}\left(\frac{4
   x_2^{16}}{x_3}-60 x_2^{15}+416 x_3 x_2^{14}-1760 x_3^2
   x_2^{13}\right.\notag\\
   &\left.+5040 x_3^3 x_2^{12}-10192 x_3^4 x_2^{11}+14560
   x_3^5 x_2^{10}-13728 x_3^6 x_2^9+5720 x_3^7 x_2^8\right)\bigg]\notag\\
   &+\left(\frac{4}{3 x_3}-\frac{2 x_1}{x_2 x_3}\right)\red{b_1}+\frac{1}{s_1^{12}}\red{b_2}\bigg[ -\frac{4 x_1^{18}}{x_2
   x_3^5}+\left(\frac{68}{x_3^5}+\frac{32}{x_3^4 x_2}\right)
   x_1^{17}+\left(-\frac{540
   x_2}{x_3^5}-\frac{400}{x_3^4}\right.\notag\\
   &\left.-\frac{36}{x_3^3 x_2}\right)
   x_1^{16}+\left(\frac{2652 x_2^2}{x_3^5}+\frac{2080
   x_2}{x_3^4}-\frac{188}{x_3^3}-\frac{632}{x_3^2 x_2}\right)
   x_1^{15}+\left(-\frac{8976 x_2^3}{x_3^5}-\frac{4992
   x_2^2}{x_3^4}\right.\notag\\
   &\left.+\frac{5720
   x_2}{x_3^3}+\frac{9072}{x_3^2}+\frac{4212}{x_3 x_2}\right)
   x_1^{14}+\left(\frac{22032 x_2^4}{x_3^5}-\frac{640
   x_2^3}{x_3^4}-\frac{40216 x_2^2}{x_3^3}-\frac{56704
   x_2}{x_3^2}\right.\notag\\
   &\left.-\frac{55380}{x_3}\right)
   x_1^{13}+\left(-\frac{39984 x_2^5}{x_3^5}+\frac{46144
   x_2^4}{x_3^4}+\frac{153384 x_2^3}{x_3^3}+\frac{204368
   x_2^2}{x_3^2}+\frac{209352 x_2}{x_3}\right.\notag\\
   &\left.+102544\right)
   x_1^{12}+\left(\frac{53040 x_2^6}{x_3^5}-\frac{174720
   x_2^5}{x_3^4}-\frac{372008 x_2^4}{x_3^3}-\frac{469664
   x_2^3}{x_3^2}-\frac{493212 x_2^2}{x_3}\right.\notag\\
   &\left.-487508 x_2\right)
   x_1^{11}+\left(-\frac{47736 x_2^7}{x_3^5}+\frac{391040
   x_2^6}{x_3^4}+\frac{597688 x_2^5}{x_3^3}+\frac{712816
   x_2^4}{x_3^2}+\frac{755856 x_2^3}{x_3}\right.\notag\\
   &\left.+755768 x_2^2+374104
   x_3 x_2\right) x_1^{10}+\left(\frac{19448
   x_2^8}{x_3^5}-\frac{613184 x_2^7}{x_3^4}-\frac{605176
   x_2^6}{x_3^3}-\frac{713856 x_2^5}{x_3^2}\right.\notag\\
   &\left.-\frac{761624
   x_2^4}{x_3}-772208 x_2^3-759032 x_3 x_2^2\right)
   x_1^9+\left(\frac{354640 x_2^8}{x_3^4}+\frac{260832
   x_2^7}{x_3^3}+\frac{482768 x_2^6}{x_3^2}\right.\notag\\
   &\left.+\frac{491560
   x_2^5}{x_3}+514576 x_2^4+504584 x_3 x_2^3+250056 x_3^2
   x_2^2\right) x_1^8+\left(-\frac{168168
   x_2^7}{x_3^2}-\frac{154336 x_2^6}{x_3}\right.\notag\\
   &\left.-216176 x_2^5-207520
   x_3 x_2^4-215360 x_3^2 x_2^3\right) x_1^7+\left(47288
   x_2^6+41360 x_3 x_2^5+65320 x_3^2 x_2^4\right.\notag\\
   &\left.+23008 x_3^3
   x_2^3\right) x_1^6+\left(7008 x_2^4 x_3^3-15408 x_2^5
   x_3^2\right) x_1^5-8808 x_2^4 x_3^4 x_1^4 \bigg]+\bigg[-\frac{4 x_1^5}{x_2^5
   x_3}\notag\\
   &+\left(\frac{16}{x_2^5}-\frac{20}{x_2^4 x_3}\right)
   x_1^4+\left(-\frac{20
   x_3}{x_2^5}+\frac{68}{x_2^4}+\frac{16}{x_2^3 x_3}\right)
   x_1^3+\left(\frac{8 x_3^2}{x_2^5}-\frac{76
   x_3}{x_2^4}-\frac{32}{x_2^3}-\frac{8}{x_2^2 x_3}\right)
   x_1^2\notag\\
   &+\left(\frac{28 x_3^2}{x_2^4}+\frac{16
   x_3}{x_2^3}+\frac{16}{x_2^2}\right) x_1-\frac{8 x_3}{x_2^2}\bigg]\red{b_3(x_3,x_1,x_2)}+\text{permutations of } x_{1,2,3}\,,
\end{align}
and 
\begin{align}
    A_{C_A}&=\frac{1}{s_1^{10}}\bigg[ \frac{2 x_1^{14}}{x_2
   x_3^4}+\left(-\frac{18}{x_3^4}-\frac{11}{2 x_3^3
   x_2}\right) x_1^{13}+\left(\frac{66
   x_2}{x_3^4}-\frac{43}{x_3^3}-\frac{1139}{18 x_3^2
   x_2}\right) x_1^{12}+\left(-\frac{114
   x_2^2}{x_3^4}\right.\notag\\
   &\left.+\frac{1223 x_2}{2 x_3^3}+\frac{88867}{90
   x_3^2}+\frac{58627}{120 x_3 x_2}\right)
   x_1^{11}+\left(\frac{42 x_2^3}{x_3^4}-\frac{2638
   x_2^2}{x_3^3}-\frac{14036 x_2}{3 x_3^2}-\frac{28521}{5
   x_3}\right) x_1^{10}\notag\\
   &+\left(\frac{198
   x_2^4}{x_3^4}+\frac{12269 x_2^3}{2 x_3^3}+\frac{468832
   x_2^2}{45 x_3^2}+\frac{142871 x_2}{10
   x_3}+\frac{1484113}{180}\right) x_1^9+\left(-\frac{374
   x_2^5}{x_3^4}\right.\notag\\
   &\left.-\frac{9029 x_2^4}{x_3^3}-\frac{216763
   x_2^3}{18 x_3^2}-\frac{80428 x_2^2}{5 x_3}-\frac{361877
   x_2}{18}\right) x_1^8+\left(\frac{198
   x_2^6}{x_3^4}+\frac{19287 x_2^5}{2 x_3^3}\right.\notag\\
   &\left.+\frac{207617
   x_2^4}{30 x_3^2}+\frac{1179863 x_2^3}{180 x_3}+\frac{253091
   x_2^2}{45}+\frac{87881 x_3 x_2}{60}\right)
   x_1^7+\left(-\frac{4674 x_2^6}{x_3^3}-\frac{23132 x_2^5}{15
   x_3^2}\right.\notag\\
   &\left.+\frac{20311 x_2^4}{30 x_3}+\frac{497777
   x_2^3}{90}+\frac{137896}{9} x_3 x_2^2\right)
   x_1^6+\left(-\frac{9592 x_2^5}{15 x_3}-\frac{124343
   x_2^4}{45}-\frac{383287}{30} x_3 x_2^3\right.\notag\\
   &\left.-\frac{26954}{5}
   x_3^2 x_2^2\right) x_1^5+\left(\frac{130303}{30} x_3
   x_2^4+\frac{133141}{45} x_3^2 x_2^3\right)
   x_1^4-\frac{122213}{180} x_2^3 x_3^3 x_1^3 \bigg]\notag\\
   &+\frac{1}{s_1^{12}}\red{\log(x_1)}\bigg[\frac{2 x_1^{16}}{x_2
   x_3^4}+\left(-\frac{20}{x_3^4}-\frac{9}{x_3^3 x_2}\right)
   x_1^{15}+\left(\frac{84
   x_2}{x_3^4}-\frac{12}{x_3^3}-\frac{157}{3 x_3^2 x_2}\right)
   x_1^{14}\notag\\
   &+\left(-\frac{180 x_2^2}{x_3^4}+\frac{587
   x_2}{x_3^3}+\frac{1129}{x_3^2}+\frac{3631}{6 x_3
   x_2}\right) x_1^{13}+\left(\frac{156
   x_2^3}{x_3^4}-\frac{3200 x_2^2}{x_3^3}-\frac{6540
   x_2}{x_3^2}\right.\notag\\
   &\left.-\frac{266383}{30 x_3}\right)
   x_1^{12}+\left(\frac{156 x_2^4}{x_3^4}+\frac{8891
   x_2^3}{x_3^3}+\frac{269164 x_2^2}{15 x_3^2}+\frac{859471
   x_2}{30 x_3}+\frac{172647}{10}\right)
   x_1^{11}\notag\\
   &+\left(-\frac{572 x_2^5}{x_3^4}-\frac{15444
   x_2^4}{x_3^3}-\frac{134847 x_2^3}{5 x_3^2}-\frac{141134
   x_2^2}{3 x_3}-\frac{193133 x_2}{3}\right)
   x_1^{10}+\left(\frac{572 x_2^6}{x_3^4}\right.\notag\\
   &\left.+\frac{18799
   x_2^5}{x_3^3}+\frac{113873 x_2^4}{5 x_3^2}+\frac{214987
   x_2^3}{5 x_3}+\frac{868598 x_2^2}{15}+\frac{146216 x_3
   x_2}{5}\right) x_1^9+\left(-\frac{18744
   x_2^6}{x_3^3}\right.\notag\\
   &\left.-\frac{53592 x_2^5}{5 x_3^2}-\frac{855203
   x_2^4}{30 x_3}-\frac{163327 x_2^3}{6}-\frac{180149}{15} x_3
   x_2^2\right) x_1^8+\left(-\frac{572
   x_2^8}{x_3^4}+\frac{18645 x_2^7}{x_3^3}\right.\notag\\
   &\left.+\frac{12528
   x_2^6}{5 x_3^2}+\frac{222251 x_2^5}{10 x_3}+\frac{46462
   x_2^4}{5}-\frac{160283}{30} x_3 x_2^3-\frac{614}{3} x_3^2
   x_2^2\right) x_1^7+\left(\frac{572
   x_2^9}{x_3^4}\right.\notag\\
   &\left.-\frac{18612 x_2^8}{x_3^3}+\frac{27927
   x_2^7}{5 x_3^2}-\frac{11636 x_2^6}{x_3}+\frac{85241
   x_2^5}{15}+2927 x_3 x_2^4-\frac{355583}{15} x_3^2
   x_2^3\right) x_1^6\notag\\
   &+\left(-\frac{156
   x_2^{10}}{x_3^4}+\frac{15345 x_2^9}{x_3^3}-\frac{270499
   x_2^8}{15 x_3^2}-\frac{5708 x_2^7}{5 x_3}-\frac{133547
   x_2^6}{15}+\frac{43481}{5} x_3 x_2^5\right.\notag\\
   &\left.+\frac{109974}{5} x_3^2
   x_2^4+\frac{7638}{5} x_3^3 x_2^3\right)
   x_1^5+\left(-\frac{156 x_2^{11}}{x_3^4}-\frac{8912
   x_2^{10}}{x_3^3}+\frac{118972 x_2^9}{5 x_3^2}-\frac{33899
   x_2^8}{10 x_3}\right.\notag\\
   &\left.-\frac{64247 x_2^7}{10}-\frac{233566}{15} x_3
   x_2^6-\frac{120734}{15} x_3^2 x_2^5+\frac{26741}{5} x_3^3
   x_2^4\right) x_1^4+\left(\frac{180
   x_2^{12}}{x_3^4}+\frac{3257 x_2^{11}}{x_3^3}\right.\notag\\
   &\left.-\frac{83028
   x_2^{10}}{5 x_3^2}+\frac{350473 x_2^9}{30 x_3}+\frac{31979
   x_2^8}{3}+\frac{22511}{10} x_3 x_2^7+\frac{24296}{5} x_3^2
   x_2^6-\frac{7638}{5} x_3^3 x_2^5\right.\notag\\
   &\left.-\frac{26741}{5} x_3^4
   x_2^4\right) x_1^3+\left(-\frac{84
   x_2^{13}}{x_3^4}-\frac{620 x_2^{12}}{x_3^3}+\frac{93641
   x_2^{11}}{15 x_3^2}-\frac{19094 x_2^{10}}{3
   x_3}-\frac{175277 x_2^9}{15}\right.\notag\\
   &\left.+\frac{111089}{15} x_3
   x_2^8+\frac{614}{3} x_3^2 x_2^7+\frac{56539}{3} x_3^3
   x_2^6-\frac{209188}{15} x_3^4 x_2^5\right)
   x_1^2+\left(\frac{20 x_2^{14}}{x_3^4}+\frac{21
   x_2^{13}}{x_3^3}\right.\notag\\
   &\left.-\frac{1113 x_2^{12}}{x_3^2}-\frac{6261
   x_2^{11}}{5 x_3}+\frac{298858
   x_2^{10}}{15}-\frac{146216}{5} x_3 x_2^9+4604 x_3^2
   x_2^8+\frac{9275}{3} x_3^3 x_2^7\right.\notag\\
   &\left.+\frac{189661}{15} x_3^4
   x_2^6-\frac{43481}{5} x_3^5 x_2^5\right) x_1-\frac{172647
   x_2^{11}}{10}+\frac{16102}{5} x_2^6 x_3^5-\frac{28677}{10}
   x_2^7 x_3^4\notag\\
   &+\frac{33123}{2} x_2^8 x_3^3-\frac{231107}{5}
   x_2^9 x_3^2+\frac{222269}{5} x_2^{10} x_3+\frac{20561
   x_2^{12}}{10 x_3}+\frac{56 x_2^{13}}{x_3^2}+\frac{8
   x_2^{14}}{x_3^3}-\frac{2 x_2^{15}}{x_3^4}\notag\\
   &+\frac{1}{x_1}\left(-\frac{3631
   x_2^{13}}{6 x_3}+\frac{20470 x_2^{12}}{3}-\frac{164381}{6}
   x_3 x_2^{11}+\frac{160228}{3} x_3^2
   x_2^{10}-\frac{328079}{6} x_3^3 x_2^9\right.\notag\\
   &\left.+\frac{95690}{3} x_3^4
   x_2^8-\frac{42167}{2} x_3^5 x_2^7+11636 x_3^6
   x_2^6\right)+\frac{1}{x_1^2}\left(\frac{157 x_2^{14}}{3 x_3}-1185
   x_2^{13}+7653 x_3 x_2^{12}\right.\notag\\
   &\left.-24187 x_3^2 x_2^{11}+43575 x_3^3
   x_2^{10}-46569 x_3^4 x_2^9+\frac{86255}{3} x_3^5 x_2^8-8091
   x_3^6 x_2^7\right)+\frac{1}{x_1^3}\bigg(\frac{9 x_2^{15}}{x_3}\notag\\
   &+4
   x_2^{14}-608 x_3 x_2^{13}+3820 x_3^2 x_2^{12}-12148 x_3^3
   x_2^{11}+24356 x_3^4 x_2^{10}-34144 x_3^5 x_2^9\notag\\
   &+37356 x_3^6
   x_2^8-18645 x_3^7 x_2^7\bigg)+\frac{1}{x_1^4}\bigg(-\frac{2
   x_2^{16}}{x_3}+22 x_2^{15}-104 x_3 x_2^{14}+264 x_3^2
   x_2^{13}\notag\\
   &-336 x_3^3 x_2^{12}+728 x_3^5 x_2^{10}-1144 x_3^6
   x_2^9+572 x_3^7 x_2^8\bigg)\bigg]+\left(\frac{x_1}{x_2 x_3}-\frac{2}{3 x_3}\right)\red{b_1}\notag\\
   &+\frac{1}{s_1^{12}}\red{b_2}\bigg[ \frac{2 x_1^{18}}{x_2
   x_3^5}+\left(-\frac{26}{x_3^5}-\frac{12}{x_3^4 x_2}\right)
   x_1^{17}+\left(\frac{150
   x_2}{x_3^5}+\frac{44}{x_3^4}-\frac{38}{x_3^3 x_2}\right)
   x_1^{16}+\left(-\frac{494 x_2^2}{x_3^5}\right.\notag\\
   &\left.+\frac{484
   x_2}{x_3^4}+\frac{1298}{x_3^3}+\frac{680}{x_3^2 x_2}\right)
   x_1^{15}+\left(\frac{968 x_2^3}{x_3^5}-\frac{4800
   x_2^2}{x_3^4}-\frac{10920
   x_2}{x_3^3}-\frac{9728}{x_3^2}-\frac{3562}{x_3 x_2}\right)
   x_1^{14}\notag\\
   &+\left(-\frac{936 x_2^4}{x_3^5}+\frac{20288
   x_2^3}{x_3^4}+\frac{45320 x_2^2}{x_3^3}+\frac{51640
   x_2}{x_3^2}+\frac{45942}{x_3}\right)
   x_1^{13}+\left(-\frac{392 x_2^5}{x_3^5}\right.\notag\\
   &\left.-\frac{52976
   x_2^4}{x_3^4}-\frac{111840 x_2^3}{x_3^3}-\frac{142016
   x_2^2}{x_3^2}-\frac{151580 x_2}{x_3}-77532\right)
   x_1^{12}+\left(\frac{2600 x_2^6}{x_3^5}\right.\notag\\
   &\left.+\frac{96096
   x_2^5}{x_3^4}+\frac{176176 x_2^4}{x_3^3}+\frac{225680
   x_2^3}{x_3^2}+\frac{269538 x_2^2}{x_3}+308338 x_2\right)
   x_1^{11}+\left(-\frac{3588 x_2^7}{x_3^5}\right.\notag\\
   &\left.-\frac{130624
   x_2^6}{x_3^4}-\frac{180648 x_2^5}{x_3^3}-\frac{215488
   x_2^4}{x_3^2}-\frac{270076 x_2^3}{x_3}-334384 x_2^2-178500
   x_3 x_2\right) x_1^{10}\notag\\
   &+\left(\frac{1716
   x_2^8}{x_3^5}+\frac{145288 x_2^7}{x_3^4}+\frac{117832
   x_2^6}{x_3^3}+\frac{136136 x_2^5}{x_3^2}+\frac{168544
   x_2^4}{x_3}+199896 x_2^3\right.\notag\\
   &\left.+190384 x_3 x_2^2\right)
   x_1^9+\left(-\frac{73788 x_2^8}{x_3^4}-\frac{37180
   x_2^7}{x_3^3}-\frac{91520 x_2^6}{x_3^2}-\frac{94408
   x_2^5}{x_3}-88660 x_2^4\right.\notag\\
   &\left.-33368 x_3 x_2^3-7484 x_3^2
   x_2^2\right) x_1^8+\left(\frac{44616
   x_2^7}{x_3^2}+\frac{38532 x_2^6}{x_3}+44616 x_2^5+3080 x_3
   x_2^4\right.\notag\\
   &\left.+16620 x_3^2 x_2^3\right) x_1^7+\left(-12272
   x_2^6-1208 x_3 x_2^5-41560 x_3^2 x_2^4-19724 x_3^3
   x_2^3\right) x_1^6\notag\\
   &+\left(16360 x_3^2 x_2^5+10096 x_3^3
   x_2^4\right) x_1^5+3956 x_2^4 x_3^4 x_1^4 \bigg]+\bigg[\frac{2 x_1^5}{x_2^5 x_3}+\frac{14 x_1^4}{x_2^4
   x_3}+\left(-\frac{6
   x_3}{x_2^5}-\frac{34}{x_2^4}\right.\notag\\
   &\left.-\frac{12}{x_2^3 x_3}\right)
   x_1^3+\left(\frac{4 x_3^2}{x_2^5}+\frac{18
   x_3}{x_2^4}+\frac{24}{x_2^3}+\frac{4}{x_2^2 x_3}\right)
   x_1^2+\left(\frac{2 x_3^2}{x_2^4}-\frac{12
   x_3}{x_2^3}-\frac{4}{x_2^2}\right) x_1\bigg]\red{b_3(x_3,x_1,x_2)}\notag\\
   &+\text{permutations of } x_{1,2,3}\,.
\end{align}
The NLP collinear function space under $\mathbb{S}_3$ permutation symmetry is the same as LP, which is composed of one logarithm $\log(x_1)$ and three transcendental weight-two functions:
\begin{equation}
    b_1=\pi^2,\quad b_2=\frac{2iD_2^{-}(z)}{s_1},\quad b_3(x_1,x_2,x_3)=\text{Li}_2\left(1-\frac{x_2}{x_1}\right)+\frac{1}{2} \log
   \left(\frac{x_1}{x_3}\right)\log
   \left(\frac{x_1}{x_2}\right)
\end{equation}
with Bloch-Wigner function $D_2^{-}(z)$ defined in Eq.~\eqref{eq:blockwigner} and $z$ introduced in Eq.~\eqref{eq:zzbardef}. 

The simplicity of the above NLP results encourages us to go to NNLP and beyond. Using the second method, we expand the integrand to N$^{10}$LP. Interestingly, we find that for $n_f$ contribution all polylogarithmic functions disappear at N${}^3$LP and beyond, which correspond to positive powers of $\lambda$ in Eq.~\eqref{eq:tripleCollinearF}. In particular, we present the first few terms:
\begin{align}
    A^{\text{N}^3\text{LP}}_{n_f}&=\frac{32 x_1}{315}+\frac{133 x_1^2+134 x_2 x_1}{1260
   x_3}+\frac{160 x_1^2 x_2-160 x_1^3}{1260 x_3^2}+\text{permutations of } x_{1,2,3} \,, \notag\\
   A^{\text{N}^4\text{LP}}_{n_f}&=\pi ^2
   \left(\frac{16 x_1 x_2}{3}-4 x_1^2\right)+\frac{5232951 x_1^2-6950496 x_1 x_2}{132300}+\frac{8050
   x_1^3+21350 x_2 x_1^2}{132300 x_3} \notag\\
   &+\frac{-12600 x_1^4+6300
   x_2 x_1^3+6300 x_2^2 x_1^2}{132300 x_3^2}+\text{permutations of } x_{1,2,3}\,, \notag\\
   A^{\text{N}^5\text{LP}}_{n_f}&=\pi ^2 \left(-12 x_1^3+\frac{56}{3} x_2
   x_1^2+\frac{16}{3} x_2 x_3 x_1\right)-\frac{2 x_1^5}{27 x_3^2}+\left(\frac{2 x_2}{135
   x_3^2}+\frac{71}{1890 x_3}\right) x_1^4\notag\\
   &+\left(\frac{8
   x_2^2}{135 x_3^2}+\frac{551 x_2}{4725
   x_3}+\frac{2090201}{17640}\right) x_1^3+\left(\frac{8
   x_2^2}{105 x_3}-\frac{6087374 x_2}{33075}\right)
   x_1^2 \nonumber \\
   & -\frac{248539 x_2 x_3 x_1}{4725}+\text{permutations of } x_{1,2,3}\,, \notag\\
   A^{\text{N}^6\text{LP}}_{n_f}&=\pi ^2
   \left(-40 x_1^4+\frac{232}{3} x_2 x_1^3-\frac{16}{3} x_2^2
   x_1^2+32 x_2 x_3 x_1^2\right)-\frac{8 x_1^6}{135 x_3^2}+\frac{29 x_1^5}{1188
   x_3}\notag\\
   &+\left(\frac{8 x_2^2}{225 x_3^2}+\frac{503 x_2}{5940
   x_3}+\frac{127687543}{323400}\right) x_1^4+\left(\frac{16
   x_2^3}{675 x_3^2}+\frac{1901 x_2^2}{14850
   x_3}-\frac{740347847 x_2}{970200}\right)
   x_1^3\notag\\
   &+\left(\frac{6395408 x_2^2}{121275}-\frac{114864394
   x_2 x_3}{363825}\right) x_1^2+\text{permutations of } x_{1,2,3}\,. 
\end{align}
Only $\pi^2$ remains at higher powers of the $n_f$ channel, while in other color channels, the polylogarithmic functions still show up. The simplicity may imply that there are some hidden symmetries, we leave it to future study.
%that it will be interesting to bootstrap collinear EEEC to higher orders.

\subsection{Squeezed limit}

Another interesting kinematic region is the squeezed limit: $x_1\to 0,x_2\sim x_3\sim\eta$ and its permutations. Using our analytic formula for EEEC, it is straightforward to extract this limit. At LP, we find,
\begin{equation}
\label{eq:squeezeExp}
\frac{1}{\sigma_\text{tot}}\frac{d^3\sigma}{dx_1 dx_2 dx_3}\overset{x_1\to 0,x_{2,3}\sim \eta}{\approx} \left(\frac{\alpha_s}{4\pi}\right)^2\frac{1}{4\pi\sqrt{-s_2^2}}\left(   \frac{B(\eta)}{x_1}  +\mathcal{O}(x_1^0) \right)\,, 
\end{equation} 
with
\begin{align}
B(\eta)&= C_F n_f T_F \left(\frac{4 \left(28
   \eta^2-82 \eta+63\right) \log (1-\eta)}{15 \eta^6}-\frac{67 \eta^3-702 \eta^2+1362 \eta-756}{45 (1-\eta)
   \eta^5}\right) \nonumber \\
   &+C_F^2 \left(-\frac{6 \left(12 \eta^3-41 \eta^2+40 \eta-9\right) \log
   (1-\eta)}{(1-\eta) \eta^6}-\frac{31 \eta^3-288 \eta^2+426 \eta-108}{2 (1-\eta) \eta^5}\right) \nonumber \\
   &+C_A C_F \left(\frac{4 \left(166 \eta^2-544 \eta+441\right) \log (1-\eta)}{15 \eta^6}-\frac{349
   \eta^3-4374 \eta^2+9174 \eta-5292}{45 (1-\eta) \eta^5}\right) \,. 
\end{align}
The functional form is similar to $\cN=4$ SYM, made of a single logarithm $\log(1-\eta)$.

\begin{figure}[!htp]
	\centering
	\begin{subfigure}{.4\linewidth}
		\includegraphics[scale=0.4]{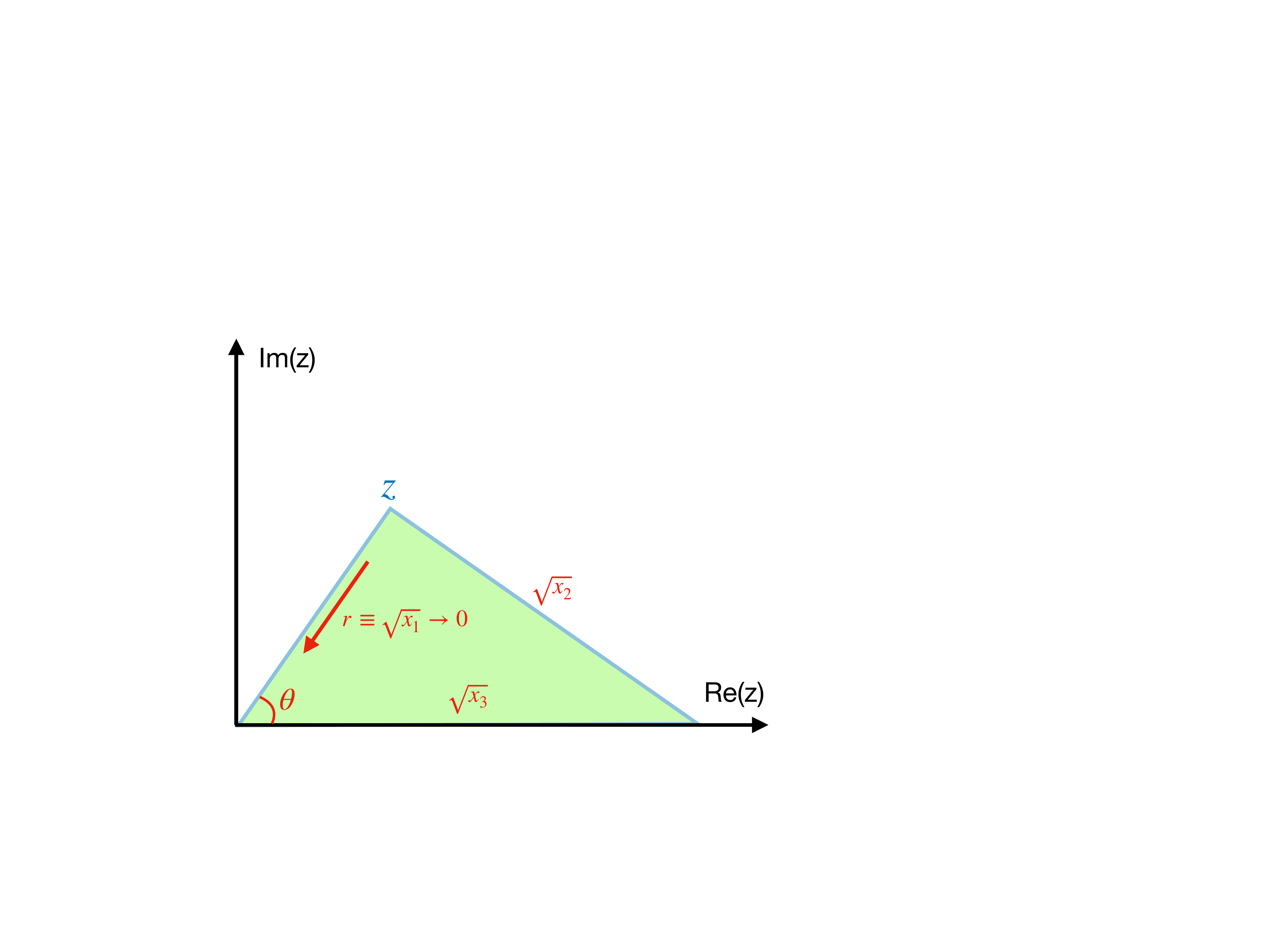}
		\caption{}
	 \label{fig:squeeze1}
	\end{subfigure}
	\qquad
	\begin{subfigure}{.4\linewidth}
		\includegraphics[scale=0.4]{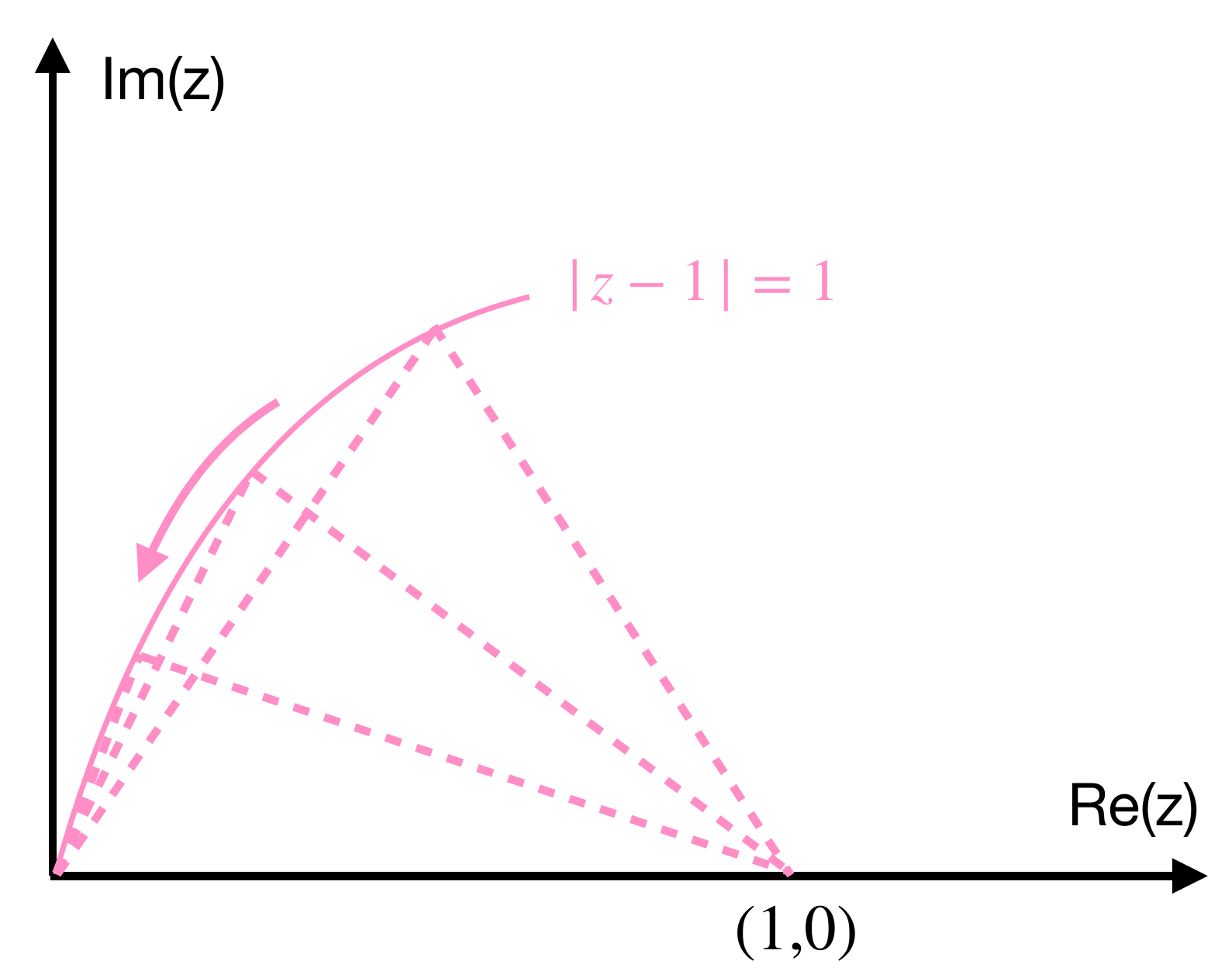}
		\caption{}
			 \label{fig:squeeze2}
	\end{subfigure}
	\caption{(a) The triangle formed by the three angular distance $\sqrt{x_1}$, $\sqrt{x_2}$ and $\sqrt{x_3}$ under the triple collinear limit. We introduce the distance $r$ from the origin to the top point and the angle $\theta$ between this side and the x-axis. (b). The squeezed limit with isosceles constraints $x_1\to 0, x_{2,3}\sim \eta$. In the triple collinear limit, this path becomes a circle $|z-1|=1$.}
	\label{fig:squeezeplot}
\end{figure}

In fact, there are ambiguities in the definition of the squeezed limit. In the above expansion, we apply the isosceles constraint first and take the zero limit of the third angular distance. However, this is not a unique choice since we can start with other configuration constraints. The ambiguity obtains a geometry interpretation if we study the squeezed limit under the triple collinear limit. As observed in Refs.~\cite{Chen:2021gdk,Chen:2020adz}, the squeezed limit is accompanied by an angular dependence. If we adopt the $z$ variable defined in Eq.~\eqref{eq:zzbardef}, one of the squeezed limits is $z\to 0$, and the expansion reads
%\begin{align}
%\label{eq:squeezeofcollinearLP}
%    \frac{1}{\sigma_\text{tot}}\frac{d^3\sigma}{dx_1 dx_2 dx_3}\approx&\left(\frac{\alpha_s}{4\pi}\right)^2\frac{1}{4\pi\sqrt{-s_2^2} x_3} \bigg[ C_F^2\left(\frac{16}{5r^2}+\frac{8(1+t^2)}{5rt }\right) \notag\\
 %   &+C_F C_A\left(\frac{5+273t^2+5t^4}{225r^2 t^2}+\frac{10+273t^2+273t^4+10t^6}{450rt^3}\right)\notag\\
 %   &+C_F T_F n_f\left(-\frac{10-39t^2+10t^4}{225r^2 t^2}+\frac{-20+39t^2+39t^4-20t^6}{450rt^3}\right)\bigg]
%\end{align}
\begin{align}
\label{eq:squeezeofcollinearLP}
    \frac{1}{\sigma_\text{tot}}\frac{d^3\sigma}{dx_1 dx_2 dx_3}\approx&\left(\frac{\alpha_s}{4\pi}\right)^2\frac{1}{4\pi\sqrt{-s_2^2} } \bigg[ \frac{C_F^2}{x_3}\left(\frac{16}{5r^2}+\frac{8(1+t^2)}{5rt }\right) \notag\\
    &+\frac{C_F C_A}{x_3}\left(\frac{5+273t^2+5t^4}{225r^2 t^2}+\frac{10+273t^2+273t^4+10t^6}{450rt^3}\right)\notag\\
    &+\frac{C_F T_F n_f}{x_3} \left(-\frac{10-39t^2+10t^4}{225r^2 t^2}+\frac{-20+39t^2+39t^4-20t^6}{450rt^3}\right)    \notag \\
    &+C_F n_f T_F \left(-\frac{24 t^4-31
   t^2+24}{225 r^2 t^2}-\frac{4 (t-1)^2 \left(t^2+1\right) (t+1)^2}{75 r
   t^3}\right) \notag \\ 
    &+C_A C_F \left(\frac{12 t^4+367 t^2+12}{225 r^2 t^2}+\frac{2 (t-1)^2
   \left(t^2+1\right) (t+1)^2}{75 r t^3}\right)  \notag \\
   & +\frac{47 C_F^2}{10 r^2}+ \mathcal{O}(x_3^0 r^0)
     \bigg]
\end{align}
with $r$ and $t=e^{i\theta}$ introduced in Fig.~\ref{fig:squeeze1}. Our definition $x_1\to 0, x_2\sim x_3\sim \eta$ then corresponds to approaching $z=0$ via the path $\theta=\frac{\pi}{2}$. In other words, the $z$ point is forced to fall into a circle located in $(1,0)$ with the radius $1$ (See Fig.~\ref{fig:squeeze2}). This gives 
\begin{equation}
\label{eq:pio2}
    \frac{1}{\sigma_\text{tot}}\frac{d^3\sigma}{dx_1 dx_2 dx_3}\approx\left(\frac{\alpha_s}{4\pi}\right)^2\frac{1}{4\pi\sqrt{-s_2^2} x_1\eta} \left( \frac{59 C_F T_F n_f}{225}+\frac{16 C_F^2}{5}+\frac{263 C_F C_A}{225} \right)+\cO(\eta^0 x_1^0)
\end{equation}
which agrees with the $\eta\to 0$ limit of $B(\eta)$. Another choice of the path is through $\theta=\frac{\pi}{4}$, as discussed in Ref.~\cite{Chen:2019bpb}, with which the expansion is different:
\begin{equation}
\label{eq:pio4}
    \frac{1}{\sigma_\text{tot}}\frac{d^3\sigma}{dx_1 dx_2 dx_3}\approx\left(\frac{\alpha_s}{4\pi}\right)^2\frac{1}{4\pi\sqrt{-s_2^2} x_1\eta} \left( \frac{13 C_F T_F n_f}{75}+\frac{16 C_F^2}{5}+\frac{91 C_F C_A}{75} \right)+\cO\left(\eta^{-1} x_1^{-1/2}\right)\,. 
\end{equation}
Interestingly, we get identical results for Eq.~\eqref{eq:pio2} and Eq.~\eqref{eq:pio4} if we take the $\mathcal{N}=1$ SYM limit by setting $T_F = 1/2, n_f=N_c, C_F =N_c, C_A = N_c$. It can be explained by looking at Eq.~\eqref{eq:squeezeofcollinearLP}, there the expression becomes $t$-independent for the squeezed limit at LP, i.e., the coefficient of $1/r^2$.

%Similarly, we can also extract the squeeze limit of the NLP triple collinear limit. The structure of the result is similar to Eq.~\eqref{eq:squeezeofcollinearLP}, given by
%\begin{align}
%    &C_A C_F \left(\frac{12 t^4+367 t^2+12}{225 r^2 t^2}+\frac{2 (t-1)^2
%   \left(t^2+1\right) (t+1)^2}{75 r t^3}\right) \nonumber \\
%   &+C_F n_f T_F \left(-\frac{24 t^4-31
%  t^2+24}{225 r^2 t^2}-\frac{4 (t-1)^2 \left(t^2+1\right) (t+1)^2}{75 r
%   t^3}\right)+\frac{47 C_F^2}{10 r^2}
%\end{align}
%\xyz{Insert the result here}
While extracting the higher power corrections in the squeezed limit is pretty straightforward once a unique definition is given, the geometry interpretation is invalid beyond LP. Nevertheless, our result indicates studying the overlap among kinematic limits themselves (triple collinear limit and squeezed limit in this case) is also theoretically important. The structure of singularities becomes clear in such a joint kinematic limit and they will be useful in investigating jet substructure. It is also interesting to ask how we can organize the power corrections under joint kinematic limits in general. We leave these possible directions for future studies.

%Taking the collinear limit,
%\begin{multline}
% B(\eta) \overset{\eta \to 0 }{\approx}   \frac{\frac{263 C_A C_F}{225}+\frac{59}{225} C_F n_f
%   T_F+\frac{16 C_F^2}{5}}{\eta }+\frac{343 C_A
%   C_F}{225}+\frac{79}{225} C_F n_f T_F+\frac{47 C_F^2}{10} + \mathcal{O}(\eta^1)
%\end{multline}

\section{Summary}\label{sec:summary}

The energy correlator is one of the most important event shape observables widely used in both precision QCD and collider physics. Proposed in the 1970s, EEC has been playing an important role in various aspects of QCD measurements and jet physics studies, such as the precise measurement of strong coupling $\alpha_s$. Three-point energy correlator, which captures more information about the scattering events, can be more powerful for probing jet substructure. 

In this paper, we calculate the three-point energy correlator at leading order in electron-positron collisions and initialize the studies of EEEC kinematics. Instead of rewriting measurement functions as cut propagators and using IBP reduction, we approach the calculation with direct phase space integration. With appropriate parameterization of the phase space $dPS_4$ and kinematic space $x_{1,2,3}$, we factorize out the Heaviside $\Theta$ function from the integral and rationalize all square roots, which allows performing the remaining integration. The QCD result is very similar to $\cN=4$ SYM EEEC, in the sense that they share the same function space that is composed of polylogarithmic functions up to transcendental weight-two.

The simplification of the leading order EEEC involves two steps. Since \texttt{HyperInt} expresses the result in terms of GPLs, we need to convert it into polylogarithms and modify their arguments with dilogarithm identities, in order to meet \texttt{Mathematica}'s branch prescription. Then we construct the raw function space by collecting rational coefficients and map it to $\cN=4$ SYM EEEC function space in Ref.~\cite{Yan:2022cye}. The linear relations between these two function spaces allow us to reduce the leading order EEEC in terms of the latter function space. With the simple function space as well as simplified rational coefficients, the file size of our analytic formula is small and the numerical evaluation is very fast. The simplicity strongly encourages us to analytically compute EEEC for gluon-initiated or $b\bar{b}$-initiated Higgs decays in the near future.

Given the multiple angular distance dependence, the EEEC kinematic space becomes more interesting. Various kinematic limits remain unexplored. In Section~\ref{sec:analysis}, we discuss the equilateral limit $x_1=x_2=x_3=x$, triple collinear limit $x_1\sim x_2 \sim x_3 \to 0$ and squeezed limit $x_1\sim 0, x_2\sim x_3\sim \eta$, and the analytic results in all limits become very simple. Under equilateral limit, the angular distance $x$ is cutoff at $x=\frac{3}{4}$, which corresponds to the coplanar configuration. Regarding the triple collinear limit, we present a method that allows us to directly compute the next-to-leading power corrections from expanding the EEEC integrand. The NLP result is simple and shares the same collinear function space as the LP. We also discuss the overlap between the triple collienar limit and the squeezed limit, where the ambiguity of the squeezed limit definition receives a geometry interpretation.
%, whose simplicity motivates one to bootstrap collinear EEEC to higher orders. 

In fact, EEEC provides a large playground for studying factorization as well as its violation for different configurations. Using our analytic result, it is straightforward to extract the needed ingredients like jet functions. Deriving the factorization theorem for specific limits and performing resummation for EEEC could be theoretically interesting and phenomenologically important. The simple mathematical structure also makes EEEC a good candidate for understanding NLP corrections and beyond. While the resummation in the triple collinear limit is in progress, the equilateral limit could be a window to investigate symmetric trijet events in $e^+ e^-$ collisions. Furthermore, all these future directions can be generalized to the analysis at hadron colliders and provide new opportunities for studying Higgs phenomenology and top physics.

\begin{acknowledgments}
We are grateful to Ian Moult, Kai Yan, and Hua Xing Zhu for their generous help and collaboration at the initial stages of this project. We thank Arindam Bhattacharya, Hao Chen, Xugan Chen, Wen Chen, Aur\'elien Dersy, Thomas Gehrmann, Yibei Li, Andreas von Manteuffel, Matthew Schwartz, Vladyslav Shtabovenko, Zhen Xu, and Yu Jiao Zhu for enlightening discussions on both the calculation and physics applications of energy correlators. We also thank Zhejiang University for its hospitality while most of this work was performed. 
T.-Z. Yang was supported by the European Research Council (ERC) under the European Union's Horizon 2020 research and innovation programme grant agreement 101019620 (ERC Advanced Grant TOPUP), and by the Swiss National Science Foundation (SNF) under contract 200020-204200.
\end{acknowledgments}

\appendix

\section{Usage of ancillary files}

In the ancillary files, we provide all the results in this paper. The usage of each file is as follows.
\begin{itemize}
	\item \texttt{EEECinQCD.nb}: The main \texttt{Mathematica} notebook. We import other files in the notebook and define a set of commands to compute EEEC. Explicitly, 
	\begin{itemize}
		%\item {\green{eeecAnalytic}} gives the full analytic expression of EEEC in QCD, where we sum all the permutations.
		\item {\green{eeec[\{x1,x2,x3\}]}} gives the value of EEEC using the analytic formula. The options ``Color'' and ``Parton'' can be used to separate different color structures or partonic subprocesses.
		\item {\green{eeecNum[\{x1,x2,x3\}]}} gives the value of EEEC using the one-fold numerical integral. We provide the same ``Color'' and ``Parton'' options.
		%\item {eeecAnalyticEqu} gives the analytic expression of equilateral EEEC in QCD. Again, we have summed all the permutations.
		\item {\green{eeecEqu[x]}} gives the value of equilateral EEEC  using the analytic formula.
		\item {\green{eeecCollLP[\{x1,x2,x3\}]}} and {\green{eeecCollNLP[\{x1,x2,x3\}]}} give the value of collinear EEEC at leading power and next-to-leading power respectively.
	\end{itemize}
	\item \texttt{Numerical.wl}: The one-fold numerical integral for EEEC in QCD, where the integrand is saved in the file \texttt{EEEConefold}. The main function is{\green{eeecNum}}.
	\item \texttt{EEECanalyticfull}: The full analytic expression of EEEC in QCD. 
	\begin{itemize}
		\item {\green{eeecQCD}}: The main formula.
		\item {\green{baseRules}}: The transcendental weight-two function space.
		\item {\green{prefactor}}: The overall normalization factor.
		\item {\green{stoxRules}}: The replacement rules from $s_{1,2}$ to $x_{1,2,3}$.
	\end{itemize} 
	\item \texttt{EEECequilateral}: The analytic expression of equilateral EEEC in QCD.
	\begin{itemize}
		\item {\green{eeecQCDEqu}}: The main formula.
		\item {\green{prefactorEqu}}: The overall normalization factor.
		\item {\green{bwrep}}: The replacement rule for Bloch-Wigner functions in equilateral EEEC.
	\end{itemize}
	\item \texttt{EEECcollinearLP}: The analytic expression of collinear EEEC in QCD at LP~\cite{Chen:2019bpb}. 
	\begin{itemize}
		\item {\green{eeecQCDCollLP}}: The main formula.
		\item {\green{CollLPBaseRules}}: The transcendental weight-two collinear function space.
%		\item {\green{collinearPrefactor}}: The overall normalization factor.
		%\item {\green{stoxRules}}: The replacement rules from $s_{1,2}$ to $x_{1,2,3}$.
	\end{itemize} 
	\item \texttt{EEECcollinearNLP}: Similar to \texttt{EEECcollinearLP}.
\end{itemize}

\bibliography{eeecQCD}{}
\bibliographystyle{JHEP}

\end{document}